\documentclass[reprint, times, 12pt, floatfix, superscriptaddress]{revtex4}

\usepackage{array}
\usepackage{graphicx}
\usepackage{appendix}
\usepackage{amsmath}
\usepackage{bm}
\usepackage{color}  
\begin{document}

\title{Phase-Field Modeling of Selective Laser Brazing of Diamond Grits}

\author{Lu Li}
\affiliation{Department of Mechanical Engineering, University of Connecticut, Storrs, CT
06269, USA}

\author{Shuai Li}
\affiliation{Department of Mechanical and Energy Engineering, Southern University of Science and Technology, Shenzhen 518055, China}

\author{Bi Zhang}

\affiliation{Department of Mechanical Engineering, University of Connecticut, Storrs, CT
06269, USA}

\affiliation{Department of Mechanical and Energy Engineering, Southern University of Science and Technology, Shenzhen 518055, China}

\author{Tai-Hsi Fan\footnote{Corresponding author. E-mail:
thfan@engr.uconn.edu}} \affiliation{Department of Mechanical
Engineering, University of Connecticut, Storrs, CT 06269, USA}

\begin{abstract}
\noindent Diamond grit is widely used in cutting, grinding, and polishing tools for its superior mechanical properties and performance in machining hard materials. Selective laser brazing (SLB) of diamond grits is a new additive manufacturing technique that has great potential to fabricate the next generation of high-performance diamond tools. However, fundamental understanding and quantitative analysis for the design and tuning of the SLB process and the resulting bonding efficiency are not yet established as the process is complicated by heating, fusion, wetting, solidification, grit migration, bonding, reaction, and the interplay between these effects. We present a thermodynamically consistent  phase-field theoretical model for the prediction of melting and wetting of SLB on diamond grits using a powder-based additive manufacturing technique. The melting dynamics is driven by laser heating in a chamber filled with argon gas and is coupled with the motion of multiple three-phase contact lines. The relevant wetting dynamics, interfacial morphology, and temperature distribution are computationally resolved in a simplified 2D configuration.
\end{abstract}
\date{\today}
\maketitle

\section{Introduction}
Synthetic diamond tools have long been developed for a variety of applications in machining metallic, glass, ceramic, and composite materials~\cite{Takahashi00,Wolfgang00,Davim02,Bai02,Konstanty03,Arno06}. Having advantages of superior hardness, tensile strength, thermal conductivity, wear resistance, self-sharpening capability, and low friction and low thermal expansion coefficient, synthetic diamond grits in a metal matrix are often used in producing cutting, grinding, and polishing tools for the machining of hard materials~\cite{Tonshoff02,Konstanty11,Artini12}. Two types of brazing filler metals are widely used in brazing diamond tools, that is, copper-based medium-temperature alloys and nickel-based high-temperature materials~\cite{Chatt91,Chen96,Huang03,Lee07,Chen14,Qi17}. The former has a relatively low operating temperature and thus lower risk of graphitization and cracks due to mild residual stress, however, it suffers from lower mechanical strength and less wear resistance. The latter has a strong affinity to diamond, great chemical resistance and wearing resistance, however, nickel could catalyze the graphitization of diamond grits at high temperature. In practices, phosphorus, boron, and carbon are often alloyed with nickel to reduce the melting temperature and alleviate the graphitization problem. Chromium can also be added as an active metal to form carbide, which enhances the bonding of diamond grits to the substrate. 

The performance and service life of diamond tools are often limited by the pullout of grits during operation, which is associated with the impregnated depth of diamond grits in the metal filler as well as the wetting profile, protrusion height from the filler metal, and the cutting conditions. Electroplating and brazing are two major techniques in fabricating surface-set diamond tools. In electroplating, diamond grits are evenly covered by the plating metal as mechanical support, however, the bonding force is relatively weak as a recess cavity region often appears in the brazing material around each grit and can not guarantee the performance of the diamond tool for an aggressive cutting at the higher speed~\cite{Ismail11,James09}. In the SLB process, the filler metal is melted by heating first, providing wetting effectively to the substrate and diamond grits. As the grits are firmly embedded in the brazing alloy, higher bonding force and thus better tool life and performance are expected than those electroplated diamond tools~\cite{James09}. 

Because brazing dynamics is complicated by heating, fusion, wetting, solidification, and chemical reaction~\cite{DusanBook}, the local heat transfer rate and temperature distribution would significantly affect the formation of the intermetallic phase and residual stress~\cite{Sebas13}. Therefore, careful control and optimization of the process are essential to the success of making brazing diamond tools. With the recent advancement of additive manufacturing of metallic parts, fabrication of metal-diamond composite is further developed by using selective laser melting (SLM) and laser cladding process~\cite{Leinenbach15,Rommel16}. Selective laser brazing (SLB) of diamond grits is an additive manufacturing technique that holds promise for adjusting the temperature, degree of melting of the filler metal, brazing profile, protrusion height of the diamond grits, and to improve bonding or adhesion force, yet reducing the risk of graphitization and micro cracks. Selective laser brazing would have advantages on better control of the spatial arrangement of diamond grits and microstructure of the metal composite matrix. However, there exist no details about transient brazing dynamics and predictive modeling tools that can facilitate basic understanding and rapid design of the process.

In this article, we focus on the first approximation of the phase transition and 2D wetting dynamics during the SLB process with an assumed fixed diamond-grit configuration. As wetting dynamics is essential for a firm bonding between diamond grits and the substrate, reaction and formation of the intermetallic phase are neglected to simplify the model. The diffuse interface or phase-field method~\cite{Cahn58,Cahn61,Penrose90,Wang93,Anderson98,Sekerka11} is applied to the derivation of phase transition and transport equations in the proposed theoretical framework. The phase-field approach has the advantage of describing transient and multiphase dynamics without explicitly tracking the moving boundaries. The details of wetting dynamics are described by the multi-component Cahn-Hilliard type equation, whereas the Allan-Cahn equation is used for solid-liquid phase transition of the filler metal. The phase-field method has been developed primarily for investigating the growth kinetics, interfacial patterning, and the stability of dendritic microstructure in metallic systems~\cite{Kobayashi93,Wheeler93,Warren95,Murray95,Karma98,Boettinger02}. Recently we have extended the phase-field approach and thermal-fluid analysis to the applications in additive manufacturing~\cite{JQ18,JQ19} and biopharmaceutical processing~\cite{Fan19,JQ19_2}. In SLB additive manufacturing process, the contact line dynamics can be described by an order parameter (phase-field variable) with interfacial boundary conditions obtained from either surface energy~\cite{Villanueva09}, geometrical contact angle~\cite{Hang07}, or imposed constraint to minimize the resulting free energy~\cite{Boyer06,Boyer11,Park16}. Here we adopt the constraint approach for its completeness in describing the evolution of contact lines and the transient wetting dynamics.

\section{Theoretical Analysis}

Figure 1 shows the simplified setting of our model system with diamond grits and powders made of the filler metal, presumed alternately placed on top of the coated substrate. Upon laser heating and melting of the filler metal, wetting and spreading introduce interfacial motion and fluid flow around the diamond grits. The subsequent solidification immerses the grits and provides bonding of the grits to the substrate surface. The following assumptions are made to facilitate the theoretical analysis: \romannumeral 1) the diamond grits are assumed fixed to the same location, that is, the motion or migration of grits is neglected, \romannumeral 2) characteristic size of the Gaussian laser beam is assumed of the same order of magnitude as grits and powders, \romannumeral 3) evaporation and condensation of the filler metal are neglected, \romannumeral 4) chemical reaction and intermetallic phase formation are not included, \romannumeral 5) the latent heat, heat capacity, density, dynamic viscosity of the filler metal are assumed constant, whereas the thermal conductivity is temperature dependent, \romannumeral 6) thermal Marangoni effect along the liquid-gas interface is neglected, \romannumeral 7) the nominal size of the powders is about the same as diamond grits with a periodic spatial arrangement, and \romannumeral 8) the ambient argon gas is assumed ideal.

\begin{figure}  \label{f1}
\centerline{\includegraphics[width=3.3in]{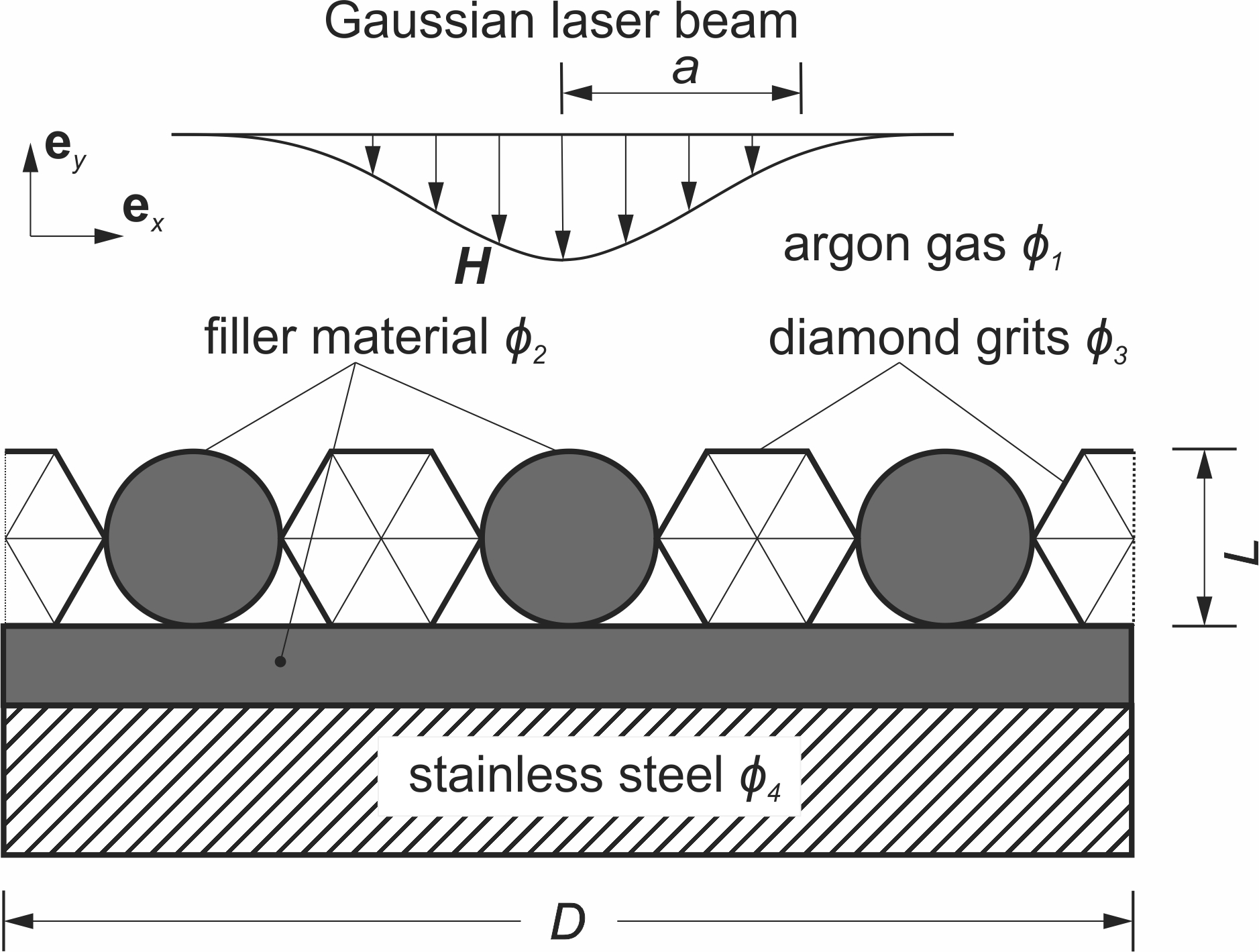}}
\caption{Schematic of selective laser brazing (SLB) process on a tool surface. A 
stainless steel substrate is coated with the filler metal, and diamond grits are affixed on top of it. Melting of the powders made of the same filler metal between grits provides bonding of grits to the substrate. $\phi_1$ to $\phi_4$ are the corresponding material volume fractions. Length $L$ is the characteristic size of grits with an assumed hexagonal shape, $D$ represents the width of the computational domain with periodic configuration, and the Gaussian beam is featured by an irradiation intensity $\textbf{\textit{H}}$ and characteristic spot radius $a$.  } 
\end{figure}

\subsection{Entropy functional}
Following the thermodynamically consistent phase-field approach~\cite{Penrose90,Wang93,Sekerka11}, we express the entropy functional of the system as
\begin{equation} \label{entropyfun}
\begin{split}
    \ \mathcal{S}'=& \int_\Omega \mathcal{L}' dV=\int_\Omega \bigg[ s \left( e,\varphi,\phi_1,\phi_2,\phi_3, \phi_4\right)\\ &~~~-\frac{1}{2}\xi_{\varphi}^2|\pmb{\nabla}\varphi|^2-\frac{1}{2}\sum_{i=1}^4\xi_i^2|\pmb{\nabla} \phi_i|^2 \bigg] dV~,
\end{split}
\end{equation}
where $\Omega$ indicates the physical and computational domain, including the substrate, filler metal, diamond grits, and the argon gas environment,  the integrand $\mathcal{L}'$ of the entropy functional includes the contributions of gradient entropy effect across the interfaces and the local entropy $s$ (per unit volume) within the bulk phase as a function of the internal energy $e$, solid-liquid phase-field variable $\varphi$, and volume fractions of the argon gas $\phi_{1}$, filler metal $\phi_{2}$, diamond $\phi_{3}$, and substrate $\phi_{4}$. The assumed constant gradient coefficients $\xi_{\varphi}$ and $\xi_{1 \sim 4}$ associated with the corresponding gradient effects are connected with the interfacial energy, thickness, as well as the constraint to the three-phase contact line. The phase-field variable $\varphi\in[-1,1]$ ($-1$ for the liquid phase and $+1$ for the solid phase) is a non-conserved order parameter to describe solid-liquid phase change, whereas $\phi_1$ to $\phi_4$ $\in[0,1]$ are material volume fractions as conserved phase-field variables under a constraint of $\sum_{i=1}^4 \phi_i=1$. The entropy functional $\mathcal{S}'$ can be modified and written as a constrained form:
\begin{equation} \label{entropyfun2}
\begin{split}
    \ \mathcal{S}=& \int_\Omega \mathcal{L} dV=\int_\Omega \bigg[ s-\lambda \left(\sum_{i=1}^4 \phi_i-1 \right)\\ &~~~-\frac{1}{2}\xi_{\varphi}^2|\pmb{\nabla}\varphi|^2 -\frac{1}{2}\sum_{i=1}^4\xi_i^2|\pmb{\nabla} \phi_i|^2 \bigg] dV~,
\end{split}
\end{equation}
where $\lambda$ is a Lagrange multiplier to be determined, and $\mathcal{L}$ is the revised integrand that incorporates the constraint. Now we consider the time derivative of the above entropy functional, 
\begin{equation} \label{varentropy}
\begin{split}
\frac{d\mathcal{S}}{dt}=\int_\Omega &\Bigg( \frac{\partial s}{\partial e}\frac{\partial e}{\partial t}+\frac{\partial s}{\partial \varphi}\frac{\partial \varphi}{\partial t}+\sum_{i=1}^4\frac{\partial s}{\partial \phi_i}\frac{\partial \phi_i}{\partial t}\\ &~~~~+\frac{\partial \mathcal{L}}{\partial \pmb{\nabla}\varphi} \pmb{\nabla}\frac{\partial \varphi}{\partial t} +\sum_{i=1}^4\frac{\partial \mathcal{L}}{\partial \pmb{\nabla}\phi_i}\pmb{\nabla}\frac{\partial \phi_i}{\partial t} \Bigg) dV~,
\end{split}
\end{equation}
which can be arranged to an Euler-Lagrange form as
\begin{equation}  \label{entropy2}
\begin{aligned}
\frac{d\mathcal{S}}{dt}=\int_\Omega &\Bigg[ \frac{\partial s}{\partial e}\frac{\partial e}{\partial t}+\left(\frac{\partial s}{\partial \varphi}-\pmb{\nabla} \cdot \frac{\partial \mathcal{L}}{\partial \pmb{\nabla}\varphi} \right)\frac{\partial \varphi}{\partial t} \\ &~+\sum_{i=1}^4\left(\frac{\partial s}{\partial \phi_i}-\pmb{\nabla} \cdot \frac{\partial \mathcal{L}}{\partial \pmb{\nabla}\phi_i} \right)\frac{\partial \phi_i}{\partial t} \\
&~+\pmb{\nabla} \cdot \left(\frac{\partial \varphi}{\partial t}\frac{\partial \mathcal{L}}{\partial \pmb{\nabla}\varphi}+\sum_{i=1}^4 \frac{\partial \phi_i}{\partial t}\frac{\partial \mathcal{L}}{\partial \pmb{\nabla}\phi_i}\right)\Bigg] dV~. \\
\end{aligned}
\end{equation}
By using variational operator $\delta$, Eq. (\ref{entropy2}) can be expressed as
\begin{equation} \label{varentropy2}
\begin{split}
\frac{d\mathcal{S}}{dt}=\int_\Omega &\Bigg[ \frac{\delta \mathcal{S}}{\delta e}\frac{\partial e}{\partial t}+\frac{\delta \mathcal{S}}{\delta \varphi}\frac{\partial \varphi}{\partial t}+\sum_{i=1}^4\frac{\delta \mathcal{S}}{\delta \phi_i}\frac{\partial \phi_i}{\partial t}\\ &~+\pmb{\nabla} \cdot \left(\frac{\partial \varphi}{\partial t}\frac{\partial \mathcal{L}}{\partial \pmb{\nabla}\varphi}+\sum_{i=1}^4 \frac{\partial \phi_i}{\partial t}\frac{\partial \mathcal{L}}{\partial \pmb{\nabla}\phi_i}\right)\Bigg] dV ~ ,
\end{split}
\end{equation}
where the first variation of the entropy functional with respect to the internal energy $e$, phase field $\varphi$, and volume fraction $\phi_i$ are
\begin{equation}   \label{se}
     \frac{\delta \mathcal{S}}{\delta e}=\frac{\partial s}{\partial e}=\frac{1}{T}~,
\end{equation}
\begin{equation}   \label{sfaisl}
 \frac{\delta \mathcal{S}}{\delta \varphi} = \frac{\partial s}{\partial \varphi} -\pmb{\nabla} \cdot \frac{\partial \mathcal{L}}{\partial \pmb{\nabla}\varphi}= \frac{\partial s}{\partial \varphi} + \xi_\varphi^2 \nabla^2 \varphi~,
\end{equation}
and
\begin{equation} \label{sfai}
 \frac{\delta \mathcal{S}}{\delta \phi_i} = \frac{\partial s}{\partial \phi_i} -\pmb{\nabla} \cdot \frac{\partial \mathcal{L}}{\partial \pmb{\nabla}\phi_i}= \frac{\partial s}{\partial \phi_i}+\xi_{i}^2 \nabla^2 \phi_i,
\end{equation}
respectively, where $T$ is temperature and $i=1$ to $4$.

Furthermore, based on the entropy transport equation, the time derivative of the entropy functional can be written as
\begin{equation} \label{Rey}
\frac{d\mathcal{S}}{dt}=\int_\Omega \left[\pmb{\nabla} \cdot \left(-\pmb{J}_s\right)+ \dot{\Gamma}+\frac{\dot{Q}}{T}  \right] dV~,
\end{equation}
where $\pmb{J}_s$ represents entropy flux, $\dot{\Gamma}$ is local entropy generation rate, which has a positive value according to the 2nd law of thermodynamics, and $\dot{Q}$ accounts for the heat source ($\dot{Q}>0$) or sink ($\dot{Q}<0$) effect. By combining Eqs. (\ref{varentropy2}) and (\ref{Rey}) and with integration over an arbitrary material domain, the differential entropy transport equation leads to
\begin{equation} \label{entropyproduction}
\begin{split}
&\pmb{\nabla} \cdot \left(-\pmb{J_s}\right)+ \dot{\Gamma}+\frac{\dot{Q}}{T} =\frac{\delta \mathcal{S}}{\delta e}\frac{\partial e}{\partial t}+\sum_{i=1}^4\frac{\delta \mathcal{S}}{\delta \phi_i}\frac{\partial \phi_i}{\partial t} \\& ~~~+\frac{\delta \mathcal{S}}{\delta \varphi}\frac{\partial \varphi}{\partial t} +\pmb{\nabla} \cdot \left(\frac{\partial \varphi}{\partial t}\frac{\partial \mathcal{L}}{\partial \pmb{\nabla}\varphi}+\sum_{i=1}^4 \frac{\partial \phi_i}{\partial t}\frac{\partial \mathcal{L}}{\partial \pmb{\nabla}\phi_i}\right),
\end{split}
\end{equation}
which forms the basis of governing transport equations and the evolution of phase-field variables that describe the laser brazing problem in hand.

\subsection{Energy equation} \label{Energy}
The differential energy equation in terms of the variational derivative of $\mathcal{S}$ can be expressed as  
\begin{equation} \label{gove}
\frac{\partial e}{\partial t}+ \pmb v \cdot\pmb{\nabla}e=- \pmb{\nabla} \cdot  \left( M_e \pmb{\nabla} \frac{\delta \mathcal{S}}{\delta e}\right)+\Phi+\dot{Q}~,
\end{equation}
and by selecting the mobility coefficient $M_e=k_TT^2$ as a function of the temperature-dependent thermal conductivity $k_T$, the first term on the right-hand side reduces to the classical Fourier heat conduction effect. The viscous dissipation function $\Phi=\pmb \sigma_\textrm{vis}:\pmb \nabla \pmb v~$ is for an assumed Newtonian fluid, with $\pmb \sigma_\textrm{vis}$ and $\pmb v$ representing the fluid flow viscous stress and the velocity field, respectively. The heat source term $\dot{Q}$ incorporates the radiation lose $\dot{Q}_{r}$ and laser irradiation $\dot{Q}_{ir}$ effects to be defined later on. Note that the local and convective derivatives on the left-hand side can be replaced by the substantial derivative of the internal energy, $De/Dt$. 

The additive internal energy is given by $e= \sum_{i=1}^4\phi_i e_i$, where $e_1, ~e_3$, and $e_4$ are the internal energy of pure argon gas, diamond, and substrate, respectively, whereas $e_2$ is the internal energy of the filler metal that follows
\begin{equation} \label{esl}
 e_2(T,\varphi)=  e_2^{(s)}(T)+P(\varphi)L_a~,
\end{equation}
where $e_2^{(s)}$ indicates internal energy density (per unit volume) of the solid phase of the filler metal, $L_a$ is an assumed constant latent heat of melting for the filler metal, and $P(\varphi)$ is an interpolation function across solid and liquid phases, here defined as
\begin{equation} \label{Pfun}
P(\varphi)=1/2-1/16\left(3\varphi^5-10\varphi^3+15\varphi\right)~.
\end{equation}  
The above polynomial function satisfies $P'=P''=0$ at $\varphi=\pm 1$~\cite{Wang93}, so that $P(1)=0$ indicates the solid phase and $P(-1)=1$ for the liquid phase. By incorporating the latent heat effect into the phase-field approach, the time derivative of internal energy can be approximated by
\begin{equation} \label{det}
\frac{D e}{D t}\simeq\sum_{i=1}^4\phi_i \frac{D e_i}{D t}=\phi_2P' L_a \frac{D \varphi}{D t}+\sum_{i=1}^4 \left( \phi_i \rho_i c_{p_i} \frac{D T}{D t} \right).
\end{equation}
We further assume that all specific heats, denoted by $c_{p_i}$, are temperature-independent, and for the solid and liquid phases of the filler metal we have $c_{{p_2}}^{(s)} \simeq c_{p_2}^{(\ell)} = c_{p_2}$. As a result, the energy equation (\ref{gove}) can be written as
\begin{equation} \label{gt}
\begin{split}
\sum_{i=1}^4 \phi_i \rho_i c_{p_i} \frac{D T}{D t}  
& =\pmb{\nabla} \cdot (k_T \pmb{\nabla} T)+\pmb \sigma_\textrm{vis}:\pmb{\nabla} \pmb v
\\ &~~~+\dot{Q}_{r}+\dot{Q}_{ir}-\phi_2P' L_a \frac{D \varphi}{D t}~,
\end{split}
\end{equation}
where temperature-dependent thermal conductivity $k_T$ has covered the contribution from each phase and can be calculated by
\begin{equation}
k_T=\sum_{i=1}^4\phi_i k_{T_i}~,
\end{equation}
with $k_{T_1}$ to $k_{T_4}$  indicating the temperature-dependent thermal conductivity for argon, filler metal, diamond grits, and the substrate, respectively. The radiation heat loss to the environment and irradiation of the laser beam at the surface $(\partial\Omega)$ of the diamond grits and filler metal are calculated by
\begin{equation} \label{heatsource1}
\dot{Q}_{r}(\textbf{x}\in\partial\Omega)=-\frac{\epsilon \sigma_{\mbox{\tiny$B$}} (T^4-T_a^4)}{W}~,
\end{equation}
and
\begin{equation} \label{heatsource2}
\dot{Q}_{ir}(\textbf{x}\in\partial\Omega)=-\frac{\alpha \pmb H \cdot \pmb n}{W}~,
\end{equation}
respectively, where $\epsilon=\sum_{i=2}^4\phi_i\epsilon_i $ is the emissivity of the surface with apparent characteristic width $W$, $\epsilon_i$ represents emissivity of each corresponding phase, $\sigma_{\mbox{\tiny$B$}}$ is the Stefan-Boltzmann constant, $T_a$ is the ambient temperature, $\alpha$ is the absorptivity of the system assumed approximately the same as $\epsilon$, $\pmb{n}$ is the outward surface normal pointing from the filler metal or diamond grits to the ambient gas environment, determined by $\pmb{n}= \pmb{\nabla }\phi_1/|\pmb{\nabla} \phi_1|$, and $\pmb H$ is the intensity of an assumed 2D Gaussian laser beam. Note that gas participation in thermal radiation is neglected here. The heat flux of the Gaussian laser beam can be approximated by
\begin{equation}
\pmb H=\frac{-\sqrt{2/\pi}\mathcal{Q}}{a} \textrm{exp} \left[ \frac{-2(x-x_0-U_at)^2}{a^2} \right] {\rm{\hat{\rm{}\textbf{e}}}_y}~,
\end{equation}
where $\mathcal{Q}$ is the laser power per unit length, $a$ is the characteristic spot radius, $x$ is horizontal coordinate, and $x_0$ is the initial position and $U_a$ is the scanning speed of the laser beam traveling along the horizontal direction ($\hat{\textbf{e}}_x$, Fig. 1). Note that the heat flux for a uniform laser beam in the test case is approximated by $\pmb H_u=-(\mathcal{Q}/D)~\hat{\textbf{e}}_y$.

\subsection{Phase-field evolution equations}
Following the entropy transport equation (Eq. \ref{entropyproduction}) and with a positive entropy generation rate $\dot{\Gamma}$, the time evolution of the non-equilibrium solid-liquid phase field $\varphi$ is assumed linearly proportional to the entropy driving force $\delta \mathcal{S}/\delta \varphi$, written as
\begin{equation} \label{csl}
\frac{\partial \varphi}{\partial t} =M_{\varphi}\frac{\delta \mathcal{S}}{\delta \varphi}=M_{\varphi} \left( \frac{\partial s}{\partial \varphi} + \xi_{\varphi}^2 \nabla^2 \varphi \right)~,
\end{equation}
where the assumed positive proportional constant $M_{\varphi}$ is the so-called interfacial mobility, and the first variation of entropy functional comes from Eq. (\ref{sfaisl}). Furthermore, the transient evolution of each volume fraction $\phi_i$ as a conserved phase-field variable follows the Cahn-Hilliard type evolution equation~\cite{Cahn58}, expressed as
\begin{equation} \label{cfai}
\begin{split}
\frac{\partial \phi_i}{\partial t} &=-\pmb{\nabla} \cdot \left[M_i \pmb{\nabla} \left(\frac{\delta \mathcal{S}}{\delta \phi_i} \right) \right]
\\ &=-\pmb{\nabla} \cdot \left [M_i \pmb{\nabla} \left( \frac{\partial s}{\partial \phi_i} + \xi_i^2 \nabla^2 \phi_i \right) \right]~
\end{split}
\end{equation}
for $i=1~\textrm{to}~4$, where the positive mobility coefficients $M_1$ to $M_4$ are for argon gas, filler metal, diamond grits, and the substrate, respectively. As mixing is avoided across the boundaries of each component in this case, the mobility coefficients are no longer associated with Fickian-type species diffusivity. Here the coefficients are determined by scaling analysis based on the comparison of characteristic time scales.  

Now to associate $\partial s/\partial \varphi$ and $\partial s/\partial\phi_i$ in Eqs. (\ref{csl}) and (\ref{cfai}) with internal energy and free energy density, the total derivative of internal energy $e(s,\varphi,\phi_1,\phi_2,\phi_3,\phi_4)$ is expressed as 
\begin{equation} \label{de}
de=Tds+\frac{\partial e}{\partial \varphi} d\varphi+\sum_{i=1}^4 \frac{\partial e}{\partial \phi_i} d\phi_i~,
\end{equation}
and thus
\begin{equation} \label{ds}
ds=\frac{1}{T}de-\frac{1}{T} \frac{\partial e}{\partial \varphi} d\varphi-\frac{1}{T}\sum_{i=1}^4 \frac{\partial e}{\partial \phi_i} d\phi_i ~.
\end{equation}
By comparing the partial derivatives of entropy $s(e,\varphi,\phi_1,\phi_2,\phi_3,\phi_4)$ with the above expression, one can establish the following relations:
\begin{equation} \label{stoevarphi}
\frac{\partial s}{\partial \varphi} \bigg)_{e,\phi_1,\phi_2,\phi_3,\phi_4}=-\frac{1}{T}\frac{\partial e}{\partial \varphi}\bigg)_{s,\phi_1,\phi_2,\phi_3,\phi_4}~,
\end{equation}
and
\begin{equation} \label{stoephi}
\frac{\partial s}{\partial \phi_i} \bigg)_{e,\varphi,\phi_{j(j \neq i)}}=-\frac{1}{T}\frac{\partial e}{\partial \phi_i}\bigg)_{s,\varphi,\phi_{j(j \neq i)}}
\end{equation}
for $i=1$ to 4. Moreover, since the Helmholtz free energy density is introduced as $f(T, \varphi, \phi_1, \phi_2, \phi_3, \phi_4)=e-Ts$, the total derivative of free energy is
\begin{equation} \label{df2}
df=d(e-Ts)=-sdT+\frac{\partial e}{\partial \varphi} d\varphi+\sum_{i=1}^4 \frac{\partial e}{\partial \phi_i} d\phi_i~, 
\end{equation}
therefore,
\begin{equation} \label{etofvarphi}
\frac{\partial e}{\partial \varphi}  \bigg)_{s,\phi_1,\phi_2,\phi_3,\phi_4}=\frac{\partial f}{\partial \varphi}  \bigg)_{T,\phi_1,\phi_2,\phi_3,\phi_4}~,
\end{equation}
and
\begin{equation} \label{etofphi}
\frac{\partial e}{\partial \phi_i} \bigg)_{s,\varphi,\phi_{j(j \neq i)}}=\frac{\partial f}{\partial \phi_i} \bigg)_{T,\varphi,\phi_{j(j \neq i)}}
\end{equation}
for $i=1$ to 4. By further incorporating the constraint of volume fraction and the Lagrange multiplier, the free energy density that is complementary to the bulk entropy density appeared in Eq. (\ref{entropyfun2}) can be formulated as
\begin{equation} \label{bulkfreeenergy}
f=\sum_{i=1}^4\phi_i f_i+f_{\rm{mix}}+T\lambda \left( \sum_{i=1}^4 \phi_i -1 \right)~,
\end{equation}
where the free energy density $f_2$ is for the filler metal, including both solid and liquid phases with an equilibrium free energy profile determined by a double-well potential~\cite{Wang93}, written as
\begin{equation} \label{slfreeenergy}
\ f_2=T \bigg[-\int_{T_m}^T \frac{e_2(T',\varphi)}{T'^2}dT'+\frac{1}{4}h_{\varphi}\left(1-\varphi^2\right)^2 \bigg],
\end{equation}
where $T_m$ is the equilibrium melting temperature of the filler metal, internal energy $e_2$ is defined in Eq. (\ref{esl}), and $h_\varphi$ is the corresponding energy barrier (per degree Kelvin) across solid and liquid phases of the filler metal. Additional energy terms $f_1$, $f_3$, and $f_4$ are the free energy densities of pure argon, diamond, and the substrate, respectively. To avoid mixing of different components, here we introduce mixing free energy $f_{\rm{mix}}$ using a double-well type potential to accommodate the enthalpy effect~\cite{JQ19_2}:
\begin{equation} \label{mixfreeenergy}
f_{\rm{mix}}=T \sum_{i=1}^4 \left[ h_i \phi_i^2(1-\phi_i)^2 \right]~,
\end{equation}
where $h_1$ to $h_4$ are the energy barriers for mixing different components. The last term on the right-hand side of Eq. (\ref{bulkfreeenergy}) takes Lagrange multiplier into account for the constraint $\sum_{i=1}^4 \phi_i=1$.

Now by combining the thermodynamic relationships above, the $\varphi$-derivative of entropy is approximated as   
\begin{equation} \label{e2f1}
\begin{split}
\frac{\partial s}{\partial \varphi}&\simeq\phi_2\left[P'L_a\frac{T-T_m}{TT_m}+h_{\varphi} \left(\varphi-\varphi^3 \right)\right]~,
\end{split}
\end{equation}
where the latent heat is assumed temperature independent, and the $\phi_i$-derivative is approximated as 
\begin{equation} \label{e2f2} 
\begin{split}
\frac{\partial s}{\partial \phi_i}&\simeq-\frac{1}{T}\frac{\partial f_{\rm{mix}}}{\partial \phi_i}-\lambda \\ & =-2h_i\phi_i(1-\phi_i)(1-2\phi_i)-\lambda~,
\end{split}
\end{equation}
where the mixing enthalpy effect dominates the free energy expression. Finally, substituting Eq. (\ref{e2f1}) into Eq. (\ref{csl}), the $\varphi$-equation for solid-liquid phase transition becomes
\begin{equation} \label{govsl} 
\begin{split}
\frac{\partial \varphi}{\partial t}  =M_{\varphi} &\bigg[\xi_{\varphi}^2\nabla^2\varphi +\phi_2 P'L_a\frac{T-T_m}{TT_m}\\ &~~~~+\phi_2 h_{\varphi} \left(\varphi-\varphi^3 \right)\bigg]~,
\end{split}
\end{equation}
where the evolution of the phase field $\varphi$ is determined by three effects: the 2nd term on the right is the thermal driving force for solid-liquid phase transition by taking elevated temperature and latent heat into account, whereas 1st and 3rd terms indicate the balance of diffusive and double-well type phase separation effects for generating and evolving a smooth yet narrow interfacial profile. Similarly, substituting Eq. (\ref{e2f2}) to Eq. (\ref{cfai}), the volume fraction phase-field equation can be formulated as 
\begin{equation} \label{govi}
\begin{split}
\frac{\partial \phi_i}{\partial t} =\pmb{\nabla} \cdot \Bigg {\lbrace} M_i\pmb{\nabla}&\bigg[ 2h_i\phi_i(1-\phi_i)(1-2\phi_i)\\ &~~~~+\lambda-\xi_i^2\nabla^2\phi_i \bigg]\Bigg {\rbrace}
\end{split}
\end{equation}
for $i=1$ to $4$ in general. The double-well term prevents the mixing of different components, the Lagrange multiplier accounts for the constraint, and the 4th-order term takes the long-ranged effect into account, which is obtained originally from the entropy gradients. In the above phase-field evolution equations, the gradient coefficients $\xi_{\varphi}^2$ and $\xi_{i}^2$, and the energy barriers $h_{\varphi}$ and $h_i$ are associated with interfacial energy and characteristic thickness of the interface, which will be explained in the following section. Note that to accommodate the fluid flow convective effect, hereafter we replace $\partial /\partial t$ by the substantial derivative $D /D t\equiv \partial /\partial t+\pmb v \cdot \pmb{\nabla}$, where $\pmb v$ is the velocity field.

\subsection{Interfacial energy and Lagrange multiplier}
The interfacial energy $\gamma$ is associated with the excess energy due to the appearance of the interface at equilibrium and can be estimated by the 1D approximation of the phase-field profile~\cite{Cahn58}. As a result, the interfacial energy at the solid-liquid interface of the filler metal can be expressed as 
\begin{equation}
\gamma_{\varphi} =\int_{-\infty}^\infty \left[T_m\xi_{\varphi}^2|\pmb \nabla \varphi|^2\right]dx=
 \frac{2\sqrt2}{3}\frac{\xi_{\varphi}^2}{W_{\varphi}}T_{m}~,
\end{equation}
where $x$ indicates the coordinate in an assumed unbounded 1D domain, $T_m$ is the reference temperature at the melting point of the filler metal, and $W_{\varphi}$ is the characteristic thickness of interface correlated with the entropy gradient coefficient through $\xi_{\varphi}^2=h_{\varphi}W_{\varphi}^2$. Similarly, the interfacial energy across two different components can be formulated by the general form:
\begin{equation}
\gamma_{ij} =\int_{-\infty}^\infty \bigg[T_{ij}^0 \left(\xi_i^2+\xi_j^2 \right)|\pmb \nabla \phi_i|^2\bigg]dx~,
\end{equation}
where $i$ and $j$ indicate the corresponding component, and $T_{ij}^0$ is the reference temperature. Considering $\phi_i\in[0,1]$, the interfacial energy between the filler metal and argon gas environment becomes
\begin{equation} \label{interenergy}
\gamma_{12}=\frac{\sqrt2}{6}\frac{ \left(\xi_1^2+\xi_2^2 \right)}{W_{12}}T_{12}^0=\frac{\sqrt2}{6}(h_1+h_2)W_{12}T_{12}^0~,
\end{equation}
where $W_{12}$ is the characteristic thickness of the interface between argon and filler metal. Similar expressions are applied to the interfacial energy between argon and diamond grits ($\gamma_{13}$), argon and stainless steel ($\gamma_{14}$), and the filler metal and diamond grits ($\gamma_{23}$). We further assign all reference temperatures to the melting point of filler metal, $T_{ij}^0=T_m$, and apply the same characteristic thickness by letting $W_{ij}=W$. Further arrangement of the four entropy gradient coefficients for their corresponding components can be formulated by the interfacial energies as
\begin{equation} \label{entropycoe}
\begin{bmatrix}
\xi_1^2 \\ \xi_2^2 \\ \xi_3^2 \\ \xi_4^2
\end{bmatrix}
=\frac{3W}{\sqrt2 T_m}
\begin{bmatrix}
~1~&1~&-1~&0~\\
~1~&-1~&1~&0~\\
~-1~&1~&1~&0~\\
~-1~&-1~&1~&2~\\
\end{bmatrix}
\begin{bmatrix}
\gamma_{12} \\ \gamma_{13} \\ \gamma_{23} \\ \gamma_{14}
\end{bmatrix}.
\end{equation}
In the multi-component system, the energy barriers are further associated with the gradient coefficients and interfacial thickness as 
\begin{equation}
h_i+h_j=\frac{\xi_i^2+\xi_j^2}{W^2}.
\end{equation}
A reduction of the relationships leads to a decoupled form: 
\begin{equation} \label{energybarrier}
h_i=\frac{\xi_i^2}{W^2}
\end{equation}
for $i=1$ to 4.

Finally, following the derivation of Boyer et al.~\cite{Boyer06,Boyer11}, the Lagrange multiplier $\lambda$ can be determined by combining Eqs. (\ref{cfai}) and (\ref{e2f2}) and substituting into the constrain $\sum_{i=1}^4\phi_i=1$, and then taking the time derivative of the constraint at an arbitrary temperature as
\begin{equation} \label{constraint}
\begin{split}
& \frac{D}{Dt}\left( \sum_{i=1}^4\phi_i\right)=0=\nabla^2 \Bigg[ \frac{1}{T}\sum_{i=1}^4 \left(M_i\frac{\partial f_{\rm{mix}}}{\partial \phi_i} \right)\\ &~~~~~~~~~~~~~
+\lambda \left(\sum_{i=1}^4 M_i \right)-\sum_{i=1}^4\left( M_i\xi_i^2\nabla^2 \phi_i\right)\Bigg]~.
\end{split}
\end{equation}
A simplified relationship was postulated by Boyer et al.~\cite{Boyer06,Boyer11} by letting
\begin{equation} \label{constr1}
 M_1\xi_1^2= M_2\xi_2^2=M_3\xi_3^2=M_4\xi_4^2=M_0~,
\end{equation}
where $M_0$ is a constant. The last term on the right-hand side of Eq. (\ref{constraint}) vanishes due to the constraint, so that the resulting Lagrange multiplier becomes
\begin{equation} \label{constr2}
\lambda=\frac{-1}{T\sum_{i=1}^4 M_i}\left( \sum_{i=1}^4 M_i\frac{\partial f_{\rm{mix}}}{\partial \phi_i}\right)~.
\end{equation}
Note that the energy barriers $h_i$ mentioned in Eq. (\ref{energybarrier}) are used to calculate the partial derivatives of the mixing energy $f_{\textrm{mix}}$, and for computing the Lagrange multiplier.

The general formulation for solving $\phi_i$ can be simplified in our case. First of all, the volume fraction of stainless steel $\phi_4$ is defined based on a fixed configuration (Fig. 1):
\begin{equation}
\phi_4=-\frac{1}{2}\textrm{tanh}\left(\frac{|y-y_s|-d_s/2}{W}\right)+\frac{1}{2}~,
\end{equation}
where $y_s$ is the center position of stainless steel in the $y$-axis, and $d_s$ is the width. Second, the volume fraction of diamond grits $\phi_3$ is also defined using a similar hyperbolic tangent function to outline the hexagonal shape, assumed a fixed configuration. Third, the volume fraction for filler metal $\phi_2$ is solved at every time instant to reflect the wetting dynamics, and the volume fraction of argon gas is calculated by the constraint $\phi_1=1-\sum_{i=2}^4\phi_i$. Finally, the Lagrange multiplier is updated from $\phi_1$ to $\phi_4$ at each time step.
\subsection{Korteweg stress and momentum equation}
During the phase transition process, the molten filler metal is assumed as a quasi-incompressible Newtonian fluid. The flow dynamics involved is described by the continuity equation and Naiver-Stokes-Korteweg momentum equation as
\begin{equation} \label{continuity}
\pmb{\nabla} \cdot \pmb v=0~,
\end{equation}
and
\begin{equation} \label{momentum}
\rho \frac{D \pmb v}{D t}=\pmb\nabla \cdot \pmb \tau-\pmb\nabla \cdot \sum_{i=1}^4 \pmb \Pi_i~,
\end{equation}
respectively, where $\rho$ is the density that includes the contributions from all components,
\begin{equation} 
\rho=\sum_{i=1}^4 \phi_i\rho_i
\end{equation}
where $\rho_1$ to $\rho_4$ stand for the mass density of each component, $\pmb v$ is the velocity field, $\pmb \tau$ represents total viscous stress, and $\pmb \Pi_i$ indicates the Korteweg stress introduced by capillarity effect across each interface. The constitutive stress-strain rate model for a Newtonian fluid can be formulated by
\begin{equation}
\pmb \tau=-p \pmb \delta+\pmb \sigma_\textrm{vis}=-p \pmb \delta+\eta\left[ \pmb \nabla \pmb v+\left( \pmb \nabla \pmb v\right)^T  \right]~,
\end{equation}
where $p$ is pressure, $\pmb\delta$ is the identity matrix, $\pmb \sigma_\textrm{vis}$ is viscous stress, and $\eta$ is a temperature-dependent dynamic viscosity, here calculated by
\begin{equation}
\eta=\phi_2\left[P\eta_2^{(\ell)}+(1-P)\eta_2^{(s)}\right]+\sum_{i=1,i\neq2}^4\phi_i\eta_i~,
\end{equation}
with $\eta_1, \eta_3$, and $\eta_4$ representing the dynamic viscosity of argon gas, diamond grits, and stainless steel, $\eta_2^{(s)}$ and $\eta_2^{(\ell)}$ are the dynamics viscosity for solid and liquid filler metal, respectively, and $P$ is the interpolation function defined by Eq. ({\ref{Pfun}}). Furthermore, using $T_m$ as a reference temperature, the Korteweg stress according to the volume fraction can be derived~\cite{Onuki07} and expressed as
\begin{equation}
\begin{split}
\pmb \Pi_i & =- T_m\xi_i^2\left(\frac{1}{2}|\pmb \nabla \phi_i|^2+\phi_i \nabla^2\phi_i\right)\pmb \delta\\ &~~~~~~~~~+ T_m\xi_i^2(\pmb\nabla\phi_i)(\pmb\nabla\phi_i)~,
\end{split}
\end{equation}
where the isotropic part of the stress tensor above can be combined with the pressure effect in the momentum equation.

To facilitate the computation, the higher-order momentum equation can be reduced by introducing a potential form~\cite{Jacqmin99,Andrea11} through a free energy functional $\mathcal{F}$ and the corresponding non-classical chemical potential $\mu_i$ below:
\begin{equation}
\begin{split}
\mathcal{F}=\int_\Omega & \bigg[ f \left( T,\varphi,\phi_1,\phi_2,\phi_3, \phi_4\right)\\ &~~~~+\frac{1}{2}T\xi_{\varphi}^2|\pmb \nabla\varphi|^2+\frac{1}{2}\sum_{i=1}^4T\xi_i^2|\pmb \nabla \phi_i|^2 \bigg] dV~,
\end{split}
\end{equation}
and
\begin{equation}
\mu_i=\frac{\delta \mathcal{F}}{\delta \phi_i}=\frac{\partial f}{\partial \phi_i}-T\xi_i^2\nabla^2\phi_i~.
\end{equation}
Here we assume that the Korteweg effect is temperature independent and the reference temperature $T_m$ is applied to the chemical potential. 
The body force term obtained from the Korteweg stress thus can be replaced by the free energy density and the chemical potential as
\begin{equation} 
\pmb\nabla \cdot \sum_{i=1}^4  \pmb \Pi_i =\pmb \nabla \left(f- T_m\sum_{i=1}^4 \xi_i^2\phi_i \nabla^2\phi_i \right)-\sum_{i=1}^4\mu_i \pmb\nabla \phi_i~.
\end{equation}
By absorbing the first term on the right-hand side of the above equation into the pressure gradient, the momentum equation can be simplified as
\begin{equation} \label{NS}
\rho \frac{D \pmb v}{D t}=-\pmb \nabla \hat{p} +\pmb\nabla \cdot\left[\eta(\pmb\nabla \pmb v+\pmb\nabla \pmb v^T)\right]+\sum_{i=1}^4 \mu_i \pmb\nabla \phi_i ~,
\end{equation}
where the modified pressure becomes
\begin{equation}
\hat{p}=p-\sum_{i=1}^4 \left(T_m\xi_i^2\phi_i\nabla^2\phi_i\right)+f~,
\end{equation}
and the computation of chemical potential is separated from the momentum equation.

\subsection{Material properties}
As temperature variation is critical in thermal and momentum transport, we summarize the relevant properties and transport coefficients that take temperature dependency into account. The density of ideal argon gas is calculated by
\begin{equation} \label{Arrho_cp}
\rho_1=\frac{p_0M_A}{R T}~,
\end{equation}
where $p_0$ is the ambient pressure, $M_A$ is the molar mass of argon, $R$ is the universal gas constant, and the data for thermal conductivity and dynamic viscosity~\cite{Eckhard10} are correlated in terms of dimensional values in MKS units and degree Kelvin as
\begin{equation} \label{Ark}
k_{T_1}\simeq1.473\times10^{-2}+2.840\times10^{-5}T~,
\end{equation}
and
\begin{equation} \label{Areta}
\eta_1\simeq1.885\times10^{-5}+3.362\times10^{-8}T~.
\end{equation}
The above linear correlations are plotted against the temperature in Fig. 2(a) for reference.

\begin{figure}  \label{f2}
\centerline{\includegraphics[width=3.1in]{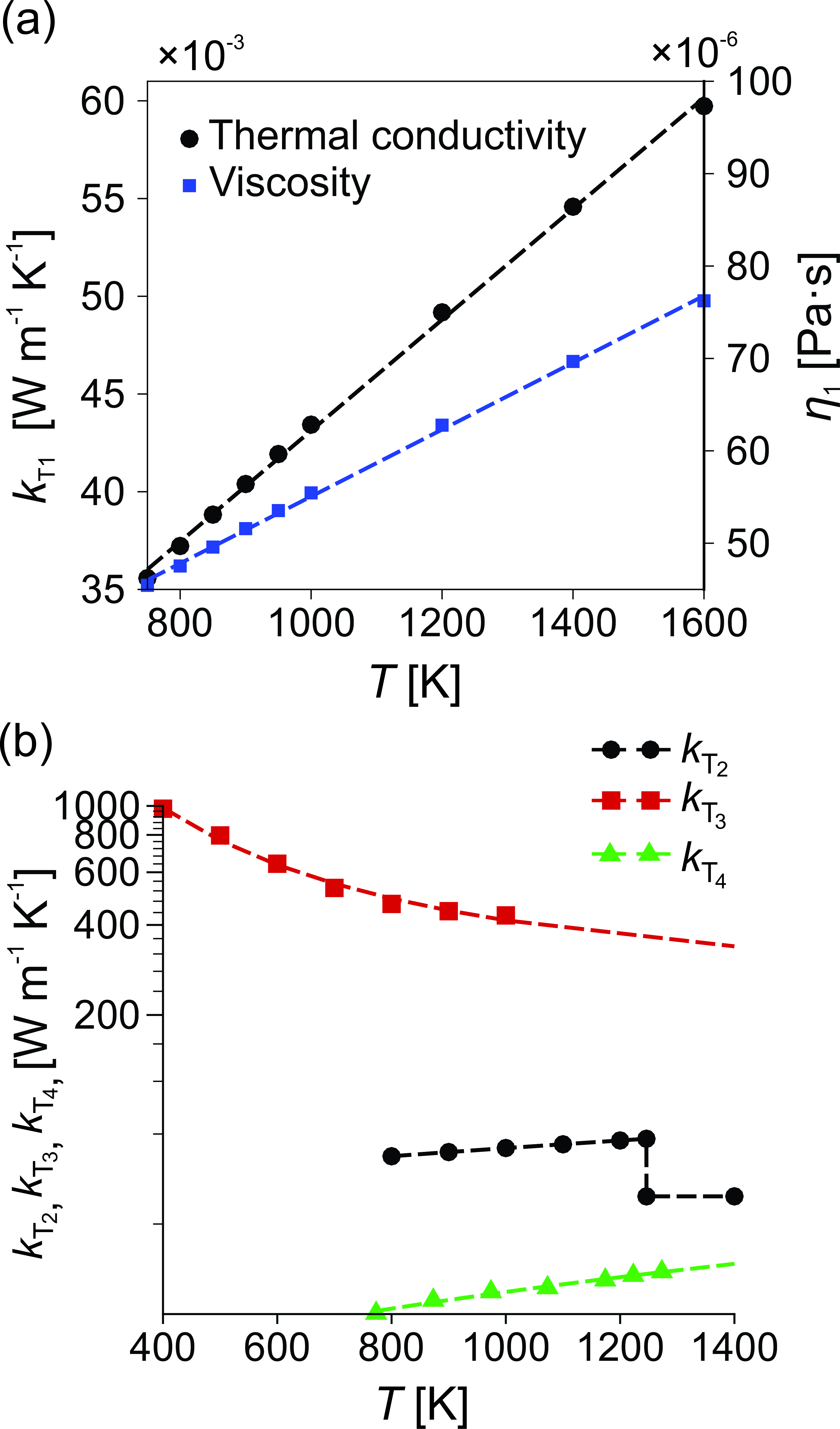}} 
\caption{(a) Temperature-dependent thermal conductivity $k_{T_1}$, and viscosity of argon gas $\eta_1$, (b) thermal conductivities of filler metal $k_{T_2}$, diamond grits $k_{T_3}$, and stainless steel substrate $k_{T_4}$.}
\end{figure}

A few material properties of commonly used nickel-based filler metals in brazing of diamond tools, such as BNi-2, BNi-3, and BNi-7 can be found in the literature~\cite{AWSA5.8,AWSBH,DusanBook}. However, these filler alloys, whether in solid or liquid form, are in general lack of temperature-dependent thermal physical properties. For the solid phase of the filler metal below melting temperature, we choose temperature-dependent thermal conductivity of a pure nickel~\cite{CRC1} instead, whereas for the molten phase we adopt a constant conductivity from a liquid nickel at its melting temperature. Across the computational domain, we utilize the interpolation function $P$, Eq. (\ref{Pfun}), to determine the overall thermal conductivity as
\begin{equation} \label{Nik1}
k_{T_2}\simeq k_{Ni}^{(s)}\left[1-P(\varphi)\right]+k_{Ni}^{(\ell)}P(\varphi)~,
\end{equation}
and $k_{Ni}^{(s)}$ represents the solid-state thermal conductivity of pure nickel and is approximated by 
\begin{equation} 
k_{Ni}^{(s)}\simeq 50.06+0.022T~,
\end{equation}
where $k_{Ni}^{(\ell)}=49.7~\rm{W}/(\rm{m} \cdot \rm{K})$ is the thermal conductivity of liquid nickel at its melting temperature~\cite{Arenas04}. The thermal conductivities of diamond~\cite{Guo10} and stainless steel~\cite{CRC2} can be estimated by
\begin{equation}
k_{T_3}\simeq227.40+7.11\times10^6T^{-1.53}~,
\end{equation}
and
\begin{equation}
k_{T_4}\simeq 9.42+0.0143T~,
\end{equation}
respectively. These conductivities are plotted in Fig. 2(b). Other material properties used in the case studies are listed in Table \ref{tab1}. A few more characteristic lengths and model parameters are included in Table \ref{tab2}.

\begin{table}[!htb]
\centering
\caption{Reference material properties.}
\label{tab1}
\begin{tabular}[t]{lr}
\hline
Parameters&Value, SI\\
\hline
mass density:&$\textrm{kg}/\textrm{m}^3$\\
$\quad \quad$filler metal $\rho_2$~\cite{CRC2}& $7810$\\
$\quad \quad$diamond $\rho_3$~\cite{CRC2}& $3500$\\
$\quad \quad$stainless steel $\rho_4$~\cite{CRC2}& $7874$\\
reference thermal conductivity $k_{T_0}$~\cite{CRC2} & $90.9$ $\textrm{W}/(\textrm{m} \cdot \textrm{K})$\\
specific heat:&$\textrm{J}/(\textrm{kg} \cdot \textrm{K})$\\
$\quad \quad$argon $c_{p_1}$~\cite{CRC2}& $520.3$\\
$\quad \quad$filler metal $c_{p_2}$~\cite{CRC2}& $490.0$ \\
$\quad \quad$diamond $c_{p_3}$~\cite{HD}& $1994.2$ \\
$\quad \quad$stainless steel $c_{p_4}$~\cite{CRC2}& $633.0$\\
dynamic viscosity of liquid filler metal $\eta_2^{(\ell)}$~\cite{Arenas04}& $0.0125$ $\textrm{Pa} \cdot \textrm{s}$ \\
interfacial energy in between:&$\textrm{J}/\textrm{m}^2$\\
$\quad \quad$solid and liquid filler metal $\gamma_{\varphi}$~\cite{Jones02} &0.347\\
$\quad \quad$nickel and argon gas $\gamma_{12}$~\cite{CRC2} &1.838\\
$\quad \quad$nickel and diamond $\gamma_{23}$~\cite{WH}&2.572\\
$\quad \quad$diamond and gas $\gamma_{13}$~\cite{WH}&3.980\\
$\quad \quad$stainless steel and argon gas $\gamma_{14}$~\cite{CRC2}&1.860\\
melting temperature of filler metal $T_m$~\cite{AWSA5.8}&$1243$ K\\
latent heat of fusion of filler metal $L_a$~\cite{CRC1} &$2.32\times 10^{9}$ $\textrm{J}/\textrm{m}^3$\\
emissivity: \\
$\quad \quad$filler metal $\epsilon_2$~\cite{MR}&$0.34$ \\
$\quad \quad$diamond grit $\epsilon_3$~\cite{Lin11} &$0.20$\\
$\quad \quad$stainless steel $\epsilon_4$~\cite{MR} &$0.40$\\
\hline
\end{tabular}
\end{table}

\begin{table}[!htb]
\centering
\caption{Additional parameters used in test cases.}
\label{tab2}
\begin{tabular}[t]{lr}
\hline
Parameters&Value, SI\\
\hline
grits size, characteristic length $L$& $2\times10^{-4}$ m\\
domain size $D=2 \pi L $&$\sim1.2\times10^{-3}$ m\\
interfacial thickness for $\varphi$ field $W_{\varphi}$&$1.6\times10^{-5}$ m\\
interfacial thickness for $\phi_i$ field $W$&$8\times10^{-6}$ m\\
temperature difference $\Delta T$&$500$ K\\
solid-liquid energy barrier $h_{\varphi}$&$18.5$ $\textrm{J}/(\textrm{m}^3\cdot\textrm{K})$\\
energy barrier for filler metal $h_2$&$91.7$ $\textrm{J}/(\textrm{m}^3\cdot\textrm{K})$\\
characteristic velocity $U$&0.73 $\textrm{m}/\textrm{s}$\\
solid-liquid interfacial mobility $M_{\varphi}$&32.1 $\textrm{m} \cdot \textrm{s} \cdot \textrm{K}/\textrm{kg}$ \\
mobility of liquid nickel $M_2$&$2.59\times 10^{-5}$ $\textrm{m}^3 \cdot \textrm{s} \cdot \textrm{K}/\textrm{kg}$ \\
2D power of laser beam $\mathcal{Q}$&$2.1\times10^5$ $\textrm{W}/\textrm{m}$\\
spot size of laser beam $a$&$100$ $\rm{\mu}\textrm{m}$\\
scanning speed of laser beam $U_a$&$0.1$ $\textrm{m}/\textrm{s}$\\
\hline
\end{tabular}
\end{table}

\subsection{Scaling and simplification}
The governing equations (\ref{gt}), (\ref{govsl}), (\ref{govi}), and (\ref{NS}), are scaled by the grit size $L$ and phase transition time scale $\tau_{\varphi}$ (Table \ref{tab3}).  The scaling and definition of reference parameters (with subscript 0) are based on filler metal as
\begin{equation}
\begin{split}
&\rho_0=\rho_2~, \quad  c_{p_0}=c_{p_2}~, \quad \text{and} \quad \eta_0=\eta_2^{(\ell)}~.
\end{split}
\end{equation}
The characteristic velocity $U$ is associated with the capillary velocity and adjusted by a constant $\beta$ (here we select $\beta=0.005$) as
\begin{equation}
U=\frac{\beta\gamma_{12}}{\eta_2^{(\ell)}}~,
\end{equation}
and the temperature is scaled by a characteristic temperature difference $\Delta T$ (assumed $500~\textrm{K}$) as
\begin{equation} 
\tilde{T}=\frac{T-T_m}{\Delta T}~.
\end{equation}
With the above reference parameters and material properties, five characteristic time scales involved in the SLB process can be determined, namely, thermal diffusion time scale $\tau_{\mbox{\tiny$T$}}$, convective time scale $\tau_c$, solid-liquid phase transition time scale $\tau_{\varphi}$, the time scale for wetting dynamics $\tau_\textrm{wet}$, and viscous diffusion time scale $\tau_\textrm{vis}$, expressed as 
\begin{equation} \label{timescale}
\begin{split}
&\tau_{\mbox{\tiny$T$}}=\frac{L^2 \rho_{0} c_{p_0}}{k_{T_0}}~, \quad \tau_c=\frac{L}{U}~,
\quad \tau_{\varphi}=\frac{1}{h_{\varphi} M_{\varphi}}~,
\\ &\tau_\textrm{wet}=\frac{L^2}{h_2 M_2}~, \quad \text{and} \quad \tau_\textrm{vis}=\frac{\rho_0 L^2}{\eta_0}~.
\end{split}
\end{equation}
The resulting values are listed in Table \ref{tab3}. The pressure and stress involved in this problem are scaled by inertia effect $\rho_0 U^2$
\begin{table}[!htb]
\centering
\caption{Time scales based on Eq. (\ref{timescale}) and relevant parameters listed in Tables \ref{tab1} $\&$ \ref{tab2}.}
\label{tab3}
\begin{tabular}[t]{lr}
\hline
Parameters&Value, s\\
\hline
thermal diffusion time $\tau_{\mbox{\tiny$T$}}$&$\sim1.68\times 10^{-3}$ \\
convective time scale $\tau_c$&$2.74\times 10^{-4}$ \\
phase transition time scale $\tau_{\varphi}$&$1.68\times 10^{-3}$ \\
wetting dynamics time scale $\tau_\textrm{wet}$&$1.68\times 10^{-5}$ \\
viscous diffusion time $\tau_\textrm{vis}$&$2.50\times 10^{-2}$ \\
\hline
\end{tabular}
\end{table}

Considering the scaling and reference parameters above, the energy equation can be simplified to a dimensionless form as
\begin{equation} \label{sgt}
\begin{split}
&\tilde{c}_p \left(\frac{\partial \tilde{T}}{\partial \tilde{t}}+\mathcal{P}_e~\tilde{\pmb v} \cdot \tilde{\nabla} \tilde{T} \right)+ \frac{\phi_2 P'}{\mathcal{S}_{te}}\left(\frac{\partial \varphi}{\partial \tilde{t}}+\mathcal{P}_e~\tilde{\pmb v} \cdot \tilde{\nabla} \varphi \right) \\&=\mathcal{L}_{e_\varphi} \tilde{\nabla} \cdot \left(\tilde{k}_T \tilde{\nabla} \tilde{T}\right)+\mathcal{B}_r \mathcal{L}_{e_\varphi} \tilde{ \pmb \sigma}_\textrm{vis}:\tilde{\nabla} \tilde{\pmb v} \\&-\frac{\alpha\mathcal{B}_{i_{ir}} \mathcal{L}_{e_\varphi}}{\mathcal{C}_{h}\tilde{a}}\sqrt{\frac{2}{\pi} } \textrm{exp} \left[ \frac{-2(\tilde{x}-\tilde{x}_0-\mathcal{P}e\tilde{U}_a\tilde{t})^2}{\tilde{a}^2} \right]\left({\hat{ \textrm{\pmb e} }_y} \cdot \pmb{n}\right)\\
& -\frac{\epsilon \mathcal{B}_{i_r} \mathcal{L}_{e_\varphi} }{\mathcal{C}_{h}}  \left[\left(1+\frac{\tilde{T} \Delta T}{T_m}\right)^4-\left(1+\frac{\tilde{T}_a \Delta T}{T_m}\right)^4\right]~, 
\end{split}
\end{equation}
where the tilde is used for scaled parameters, heat capacity $\tilde{c}_p=\sum_{i=1}^4 \phi_i \tilde{\rho_i} \tilde{c}_{p_i}$, thermal conductivity $\tilde{k}_T=\sum_{i=1}^4\phi_i \tilde{k}_{T_i}$, emissivity $\epsilon=\sum_{i=2}^4 \phi_i\epsilon_i$, absorptivity is assumed the same as emissivity, $\alpha=\epsilon$, and the dimensionless groups are defined as
\begin{equation}
\begin{split}
&\mathcal{P}_e=\frac{\tau_{\varphi}}{\tau_c}~,\quad \mathcal{S}_{te}=\frac{\rho_0c_{p_0}\Delta T}{L_a}~, \quad \mathcal{L}_{e_\varphi}=\frac{\tau_{\varphi}}{\tau_{\mbox{\tiny$T$}}}~,\\&  \mathcal{B}_r=\frac{\eta_{0}U^2}{{k_{T_{0}}} \Delta T}~,\quad \mathcal{B}_{i_r}=\frac{\sigma_{\mbox{\tiny$B$}} T_m^4 L}{{k_{T_{0}}} \Delta T}~, \\ &  \mathcal{B}_{i_{ir}}=\frac{\mathcal{Q}}{{k_{T_{0}}} \Delta T}~ \quad \textrm{and} \quad \mathcal{C}_{h}=\frac{W}{L}~. 
\end{split}
\end{equation}
The Peclet number $\mathcal{P}_e$ compares the phase transition and convective time scales, Stefan number $\mathcal{S}_{te}$ measures the ratio of sensible heat to latent heat, interfacial Lewis number $\mathcal{L}_{e_\varphi}$ measures the ratio of the phase transition to thermal diffusion time scales, Brinkman number $\mathcal{B}_r$ compares the viscous dissipation to the heat conduction effects, Biot number $\mathcal{B}_{i_{ir}}$ measures the ratio of irradiation to heat conduction effect, Biot number $\mathcal{B}_{i_r}$ measures the radiation heat transfer to heat conduction effect, and $\mathcal{C}_{h}$ is the Cahn-Hilliard number indicating the relative thickness of the interface to the length scale. The scaled $\varphi$-equation can be expressed as 
\begin{equation} \label{sgsl}
\begin{aligned}
\frac{\partial \varphi}{\partial \tilde{t}}+\mathcal{P}_e~ \tilde{\pmb v} \cdot  \tilde{\nabla} \varphi&=\mathcal{C}_{h_{\varphi}}^2 \tilde{\nabla}^2 \varphi+\phi_2(\varphi-\varphi^3)\\
&~~~~~+\phi_2 P' \Lambda_{\varphi} \frac{\tilde{T}}{1+(\Delta T/{T_m})\tilde{T}}~,
\end{aligned}
\end{equation}
where an additional Cahn-Hilliard number $\mathcal{C}_{h_{\varphi}}$ represents the thickness of solid-liquid interface to the length scale, and the phase-change number $\Lambda_{\varphi}$ describes the ratio of latent heat of fusion to the interfacial energy, defined as
\begin{equation} 
\mathcal{C}_{h_{\varphi}}=\frac{W_{\varphi}}{L}~, \quad \text{and}  \quad \Lambda_{\varphi}=\frac{L_a \Delta T}{h_{\varphi} T_m^2}~,
\end{equation}
respectively. Furthermore, the governing equation for the volume fraction of the filler metal $\phi_2$ reduces to
\begin{equation} \label{sgfai}
\begin{split}
\frac{\partial \phi_2}{\partial \tilde{t}}+\mathcal{P}_e~\tilde{\pmb v} \cdot \tilde{\nabla} \phi_2&=\frac{\tau_{\varphi}}{\tau_\textrm{wet}} \tilde{\nabla} \cdot \Bigg\lbrace \tilde{\nabla} \bigg[\frac{\lambda}{h_2}-\mathcal{C}_{h}^2 \tilde{\nabla}^2{\phi_2}\\&~~~+2\phi_2(1-\phi_2)(1-2\phi_2)\bigg]\Bigg\rbrace~.
\end{split}
\end{equation}
Finally, the scaled Navier-Stokes-Korteweg momentum equation is written as
\begin{equation} \label{sgns}
\begin{aligned}
\frac{\tilde{\rho}}{\mathcal{S}_c} \left( \frac{\partial \tilde{\pmb v}}{\partial t}+\mathcal{P}_e~\tilde{\pmb v} \cdot \tilde{\nabla} \tilde{\pmb v} \right)&\simeq-\mathcal{R}_e \nabla \tilde{p}+\mathcal{R}_e\Gamma \sum_{i=1}^4 \left( \tilde{\mu}_i\tilde{\nabla} \phi_i \right)\\ &~~~~~~~+ \tilde{\nabla} \cdot \left[\tilde{\eta}(\tilde{\nabla} \tilde{\pmb v}+ \tilde{\nabla} \tilde{\pmb v}^T)\right]~,
\end{aligned}
\end{equation}
where the scaled chemical potential $\tilde{\mu}_i=\mu_i/(T_m h_2)$, density $\tilde{\rho}=\sum_{i=1}^4 \phi_i \tilde{\rho}_i$, and the dynamic viscosity is
\begin{equation}
\tilde{\eta}=\phi_2\left[P\tilde{\eta}_2^{(\ell)}+(1-P)\tilde{\eta}_2^{(s)}\right]+\sum_{i=1,i\neq2}^4\phi_i \tilde{\eta}_i~.
\end{equation}
Note that the dynamic viscosities for all solid components are much larger than the dynamic viscosity of liquid filler metal, here we assume $\tilde{\eta}_3=\tilde{\eta}_4=\tilde{\eta}_2^{(s)}\simeq10^5$. The Schmidt number $\mathcal{S}_c$ compares the phase transition to viscous time scales, Reynolds number $\mathcal{R}_e$ indicates the inertia to viscous effects, and Korteweg number $\Gamma$ measures the ratio of energy barrier of filler metal to the kinetic energy, defined as
\begin{equation} \label{dg4}
\begin{split}
& \mathcal{S}_c=\frac{\tau_{\varphi}}{\tau_\textrm{vis}}~,\quad \mathcal{R}_e=\frac{\tau_\textrm{vis}}{\tau_c}~,
\quad \text{and} \quad \Gamma=\frac{T_m h_2}{\rho_0 U^2}~,
\end{split}
\end{equation}
respectively. The dimensionless groups and their corresponding values are listed in Table \ref{tab4} for order-of-magnitude comparison.

\begin{table}[!htb]
\centering
\caption{Dimensionless groups.}
\label{tab4}
\begin{tabular}[t]{lr}
\hline
Dimensionless group&Value\\
\hline
Peclet number $\mathcal{P}_e$&$6.14$\\
Stefan number $\mathcal{S}_{te}$&$0.82$\\
Lewis number $\mathcal{L}_{e_\varphi}$&$1$\\
Brinkman number $\mathcal{B}_r$&$1.47\times10^{-7}$\\
Biot number for radiation $\mathcal{B}_{i_r}$&$5.96\times 10^{-4}$\\
Biot number for irradiation $\mathcal{B}_{i_{ir}}$&$0.018$\\
Cahn-Hilliard number for $\varphi$ field $\mathcal{C}_{h_{\varphi}}$&$0.08$\\
phase change number $\Lambda_{\varphi}$&$4.07\times10^{4}$\\
Cahn-Hilliard number for $\phi_i$ field $\mathcal{C}_{h}$&$0.04$\\
Reynold number $\mathcal{R}_e$&$91.2$\\
Schmidt number $\mathcal{S}_c$&$0.067$\\
Korteweg number $\Gamma$&$27.4$\\
\hline
\end{tabular}
\end{table}

In summary, the governing system and the assumptions are proposed to simulate the phase transition and wetting dynamics of diamond grits brazing process using nickel-based filler metal. The computational algorithm is developed for the scaled formulation, and in general applicable to 2D and 3D cases. The fully coupled governing equations are for solving solid-liquid phase transition dynamics $\varphi$, the volume fraction of the filler metal $\phi_2$, temperature $\tilde{T}$, and velocity field $\tilde{\pmb{v}}$, along with initial and periodic boundary conditions. 

\newpage
\section{Results and Discussion}
To demonstrate the transient dynamics, numerical tests are performed by applying an alternative spatial arrangement of the filler powders and diamond grits (Fig. 1). The spatial discretization of the scaled equations is based on a uniform 2D mesh $(800\times800)$ and with a periodic boundary condition to facilitate the computation using Fourier spectral method. The thermal conductivity of the substrate material has been adjusted to provide a quasi-insulation boundary condition at the bottom of the computational domain. The pseudo-spectral scheme is used to discretize nonlinear terms. The momentum equation is solved by using the projection formulation to decouple velocity field from the pressure field, and the solid phase is simply taken as an assumed fluid with much higher (at least five orders of magnitude) viscosity than the molten filler metal. The nonlinear effects induced by variable transport coefficients are discretized using the algorithm provided by Zhu et al.~\cite{Zhu99}. The temporal discretization applies the forward Euler integration scheme with uniform time step $h=10^{-5}$ and a semi-implicit spatial discretization is applied for all test cases. The transient simulations are carried out for about $5\times10^{5}$ time steps until the formation of meniscus around the diamond grits. Note that there is no smooth or adaptive scheme applied to the moving interfaces in the phase-field approach, which can be further extended to a variety of conditions with different powder size and configuration or spatial arrangement. Next, we present a validation of three-phase contact line dynamics and the simulation of SLB using stationary and moving laser beams.  

\subsection{Free energy of a ternary system}
\begin{figure}  \label{f3}
\centerline{\includegraphics[width=3.3in]{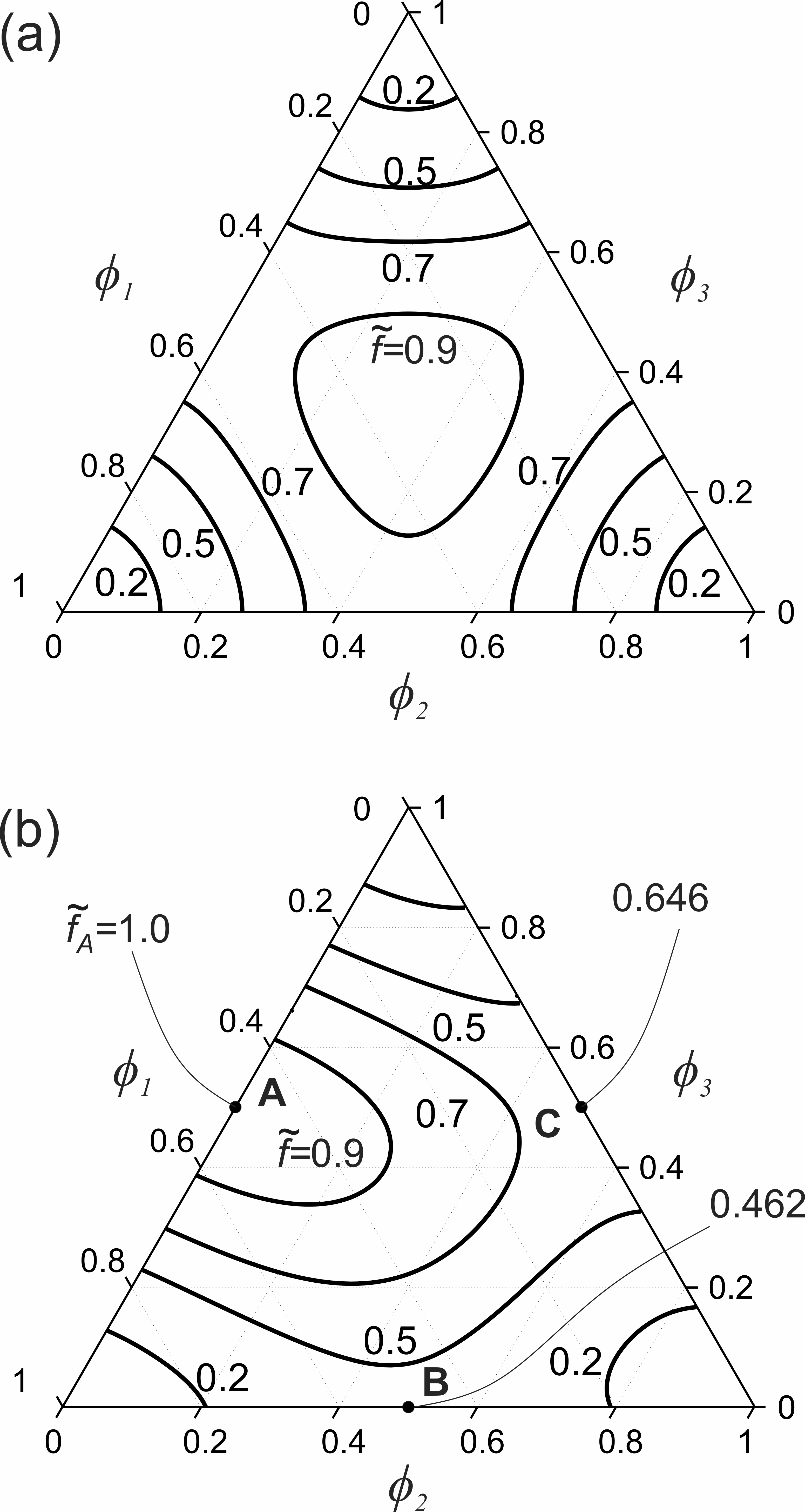}} 
\caption{Normalized free energy of an unbiased ternary system (a), and for an argon($\phi_1$)-filler metal($\phi_2$)-diamond($\phi_3$) system (b) at equilibrium.}
\end{figure}
Figure 3 demonstrates the contour map of normalized free energy and its relation with the contact angle under the steady-state condition for an assumed unbiased ternary system (Fig. 3a) and the argon-filler-diamond system (Fig. 3b), where the free energy $f_\textrm{mix}$ (Eq. \ref{mixfreeenergy}) is scaled by its maximum value. For the unbiased case, $\gamma_{12}=\gamma_{13}=\gamma_{23}$, the energy barriers are based on a similar relationship as given in Eqs. (\ref{entropycoe}) and (\ref{energybarrier}), and the result indicates that the free energy has three locally minimum points at ($\phi_1,~\phi_2, ~\phi_3$) = (1, 0, 0), (0, 1, 0), and (0, 0, 1), implying a separation of the three-component system which is equally weighted by a triple-well type energy potential with the global maximum located at the center point of the energy landscape, i.e., ($\phi_1,~\phi_2, ~\phi_3$) = (1/3, 1/3, 1/3). On the other hand, for the argon-filler-diamond system, the energy barriers are calculated from interfacial energy $\gamma_{12}$, $\gamma_{13}$, and $\gamma_{23}$ listed in Table \ref{tab2}. The free energy has three local minima at the same locations as the unbiased system with a shifted global maximum. Points A($\phi_1=0.5,~\phi_2=0, ~\phi_3=0.5$), B(0.5, 0.5, 0), and C(0, 0.5, 0.5) in Fig. 3b correspond to the interfaces between argon and diamond, argon and filler, and filler and diamond, respectively, with $\tilde{f}_A>\tilde{f}_C>\tilde{f}_B$, meaning that the system prefers to have a larger interfacial area between argon-filler or to reduce interface formation between argon and diamond. According to the definition of $f_\textrm{mix}$ and interfacial energy relationship, Eq. (\ref{interenergy}), the free energy at points A, B, and C can be correlated to interfacial free energy as $\tilde{f}_A:\tilde{f}_B:\tilde{f}_C =(h_1+h_3):(h_1+h_2):(h_2+h_3)=\gamma_{13}:\gamma_{12}:\gamma_{23}$, which is associated with the steady-state contact angle of a sessile droplet of liquid filler metal on top of a diamond plate in an argon gas environment. The contact angle is $\theta=\textrm{cos}^{-1} \left[(\tilde{f}_A-\tilde{f}_C)/\tilde{f}_B\right]\simeq40 ^\circ$. 

\begin{figure}
\centerline{\includegraphics[width=3.8in]{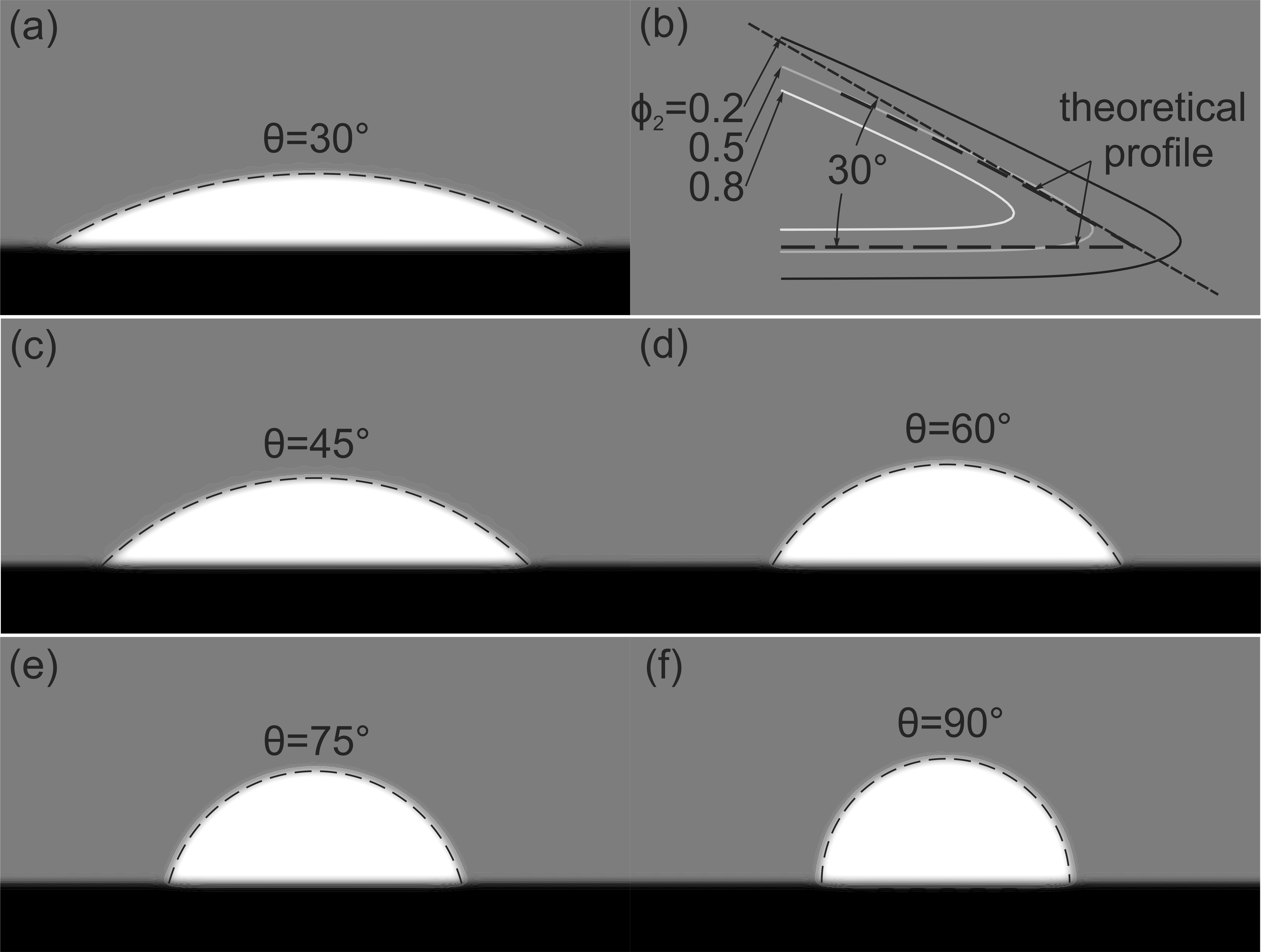}}
\caption{A sessile liquid droplet on top of a solid substrate at different static contact angles. The gravity effect is neglected, and the theoretical profiles can be well approximated by a portion of a circle for $\theta=$ $30^\circ$ (a), $45^\circ$ (c), $60^\circ$ (d), $75^\circ$ (e), and $90^\circ$ (f). The black area mimics the solid substrate ($\phi_3$), the grey area is the for the vapor phase ($\phi_1$), and the white area is the droplet ($\phi_2$). The zoom-in view (b) shows phase-field contours at $\phi_2=0.2$,~$0.5$,~$0.8$ near the tri-junction point, and the theoretical profile (long-dash line) at a contact angle of $30^\circ$. Reference parameters: $T_m=1243$ K, $W=1\times10^{-6}$ m, $\gamma_{13}=3.980$~$\rm{J}/m^2$, $\gamma_{12}=1.838$~$\rm{J}/m^2$,  and $\gamma_{23}=\gamma_{13}-\gamma_{12}\cos \theta$.}
\end{figure}

As a simple test of the steady-state model, wetting of the solid plate by a liquid droplet within a vapor environment under different contact angles is tested by evolving $\phi_2$ equation from a spherical droplet to a fully relaxed state. The steady-state result shown in Fig. 4 at different contact angles agrees well with the theoretical profiles (dashed lines).

\subsection{Brazing process using a stationary laser beam}

\begin{figure*}  \label{f5}
\centerline{\includegraphics[width=\textwidth]{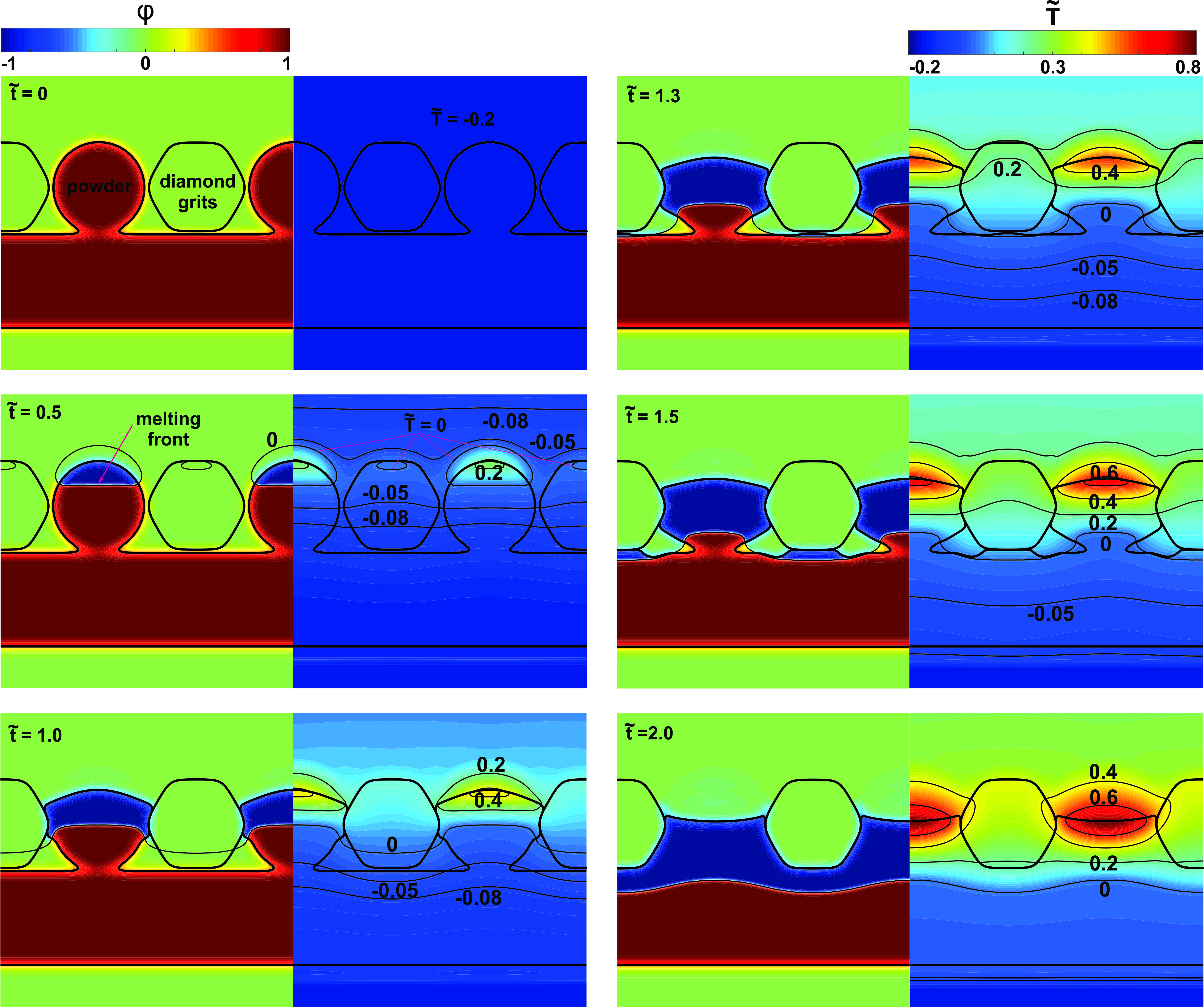}} 
\caption{Transient evolution of phase field $\varphi$ for describing solid-liquid phase change and the scaled temperature field $\tilde{T}$ driven by heating from a laser beam with uniform intensity at six scaled time instants $\tilde{t}$=0, 0.5, 1.0, 1.3, 1.5 and 2.0. The sequential plots include temperature contours (thin solid lines) and boundary profiles of metals and diamond grits located at $\phi_2=0.5$ and $\phi_3=0.5$, respectively. The time scale is defined by $\tau_{\varphi}=1.7 \times 10^{-3} \rm{s}$, and the temperature is scaled as $\tilde{T}=(T-T_m)/(\Delta T)$, with melting temperature of the filler metal $T_m=1243$ K and characteristic temperature difference $\Delta T=500$ K.}
\end{figure*}

Figure 5 demonstrates the transient dynamics of brazing of nickel-based filler metals using an assumed stationary laser beam with uniform irradiation intensity for the heating and melting process. To demonstrate the wetting process driven by interfacial energy alone, first, we neglect the convective effect in the process simulation. The 2D laser power $\mathcal{Q}$ is $2.1\times 10^5~\rm{W}/ \rm{m}$, spot radius $a\to \infty$, and the scanning speed is set to $U=0$. The initial configuration at $\tilde{t}=0$ is arranged by placing three powders of filler metals and three diamond grits of equal size and arranged alternatively on top of a layer of filler metal attached to the substrate (Fig. 1). Note that powder size $L=200~\mu$m and domain size $D\simeq1200~\mu$m. The process time scale $\tau_{\varphi}$ is about 1.7$\times$10$^{-3}$ second. The initial temperature $\tilde{T}=-0.2$ is uniform for all components including the argon gas environment. Figure 5 includes six sequential plots at scaled time instants $\tilde{t}$ = 0, 0.5, 1.0, 1.3, 1.5, and 2.0. At each time instant, a color map on the left is for the phase field $\varphi$ and side-by-side compared with a scaled temperature map $\tilde{T}$ on the right. The evolution of the filler metals and the configuration profiles of grits are indicated by thick solid lines determined by $\phi_2=0.5$ and $\phi_3=0.5$, whereas a few selected temperature contours are shown by thin solid lines for reference. In this case, the onset of melting appears at the top surface of the filler metal powders during the early stage of laser heating, shown at time instant around $\tilde{t}=0.5$. At this moment, the filler powders are not in contact with diamond grits, and thus a circular powder shape is maintained owing to a strong surface tension effect. The downward advancing of the melting front overlaps with the melting temperature contour $\tilde{T}=0$, which validates the basic assumption that the process is thermally controlled and the solid-liquid interface is near an equilibrium state. The corresponding temperature map shows that the thermal diffusion wave has penetrated the filler metals and diamond grits at $\tilde{t}=1.0$, as expected from scaling estimation. Furthermore, the evolution of the molten filler metals driven by wetting effect occurs at a later stage after time instant $\tilde{t}$ reaches about 1.0 (Fig. 5). The three-phase contact line moves upward and downward to coat the diamond grits, shown in the sequential plots at time instants $\tilde{t}=1.0, ~1.3,$ and 1.5. During the process, higher temperature appears at the top surface of the filler powders due to higher absorptivity of thermal radiation ($\alpha\simeq0.34$) compared to diamond grits ($\alpha\simeq0.2$ ). At $\tilde{t}=1.0$ and 1.3, one can observe a gradually deeper melting temperature contour appeared in the diamond grits compared with the filler metal. This is because of the absorption of latent heat into the filler metal in addition to its relatively low thermal conductivity ($k_{T_{2}}=73.5$ vs. $k_{T_{3}}=281.6~ \rm{W}\rm{m}^{-1} \rm{K}^{-1}$ at $T=T_m$, shown in Fig. 2b). Eventually,  the molten powders are fused with coated layer on the steel substrate, and the evolving interface fills the gap region as shown at $\tilde{t}=2.0$. At this moment, a meniscus nearly at equilibrium is formed between diamond grits in order to provide the required bonding force. Continuous heating leads to further melting of the coated material.  

\begin{figure}  \label{f6}
\centerline{\includegraphics[width=3.8in]{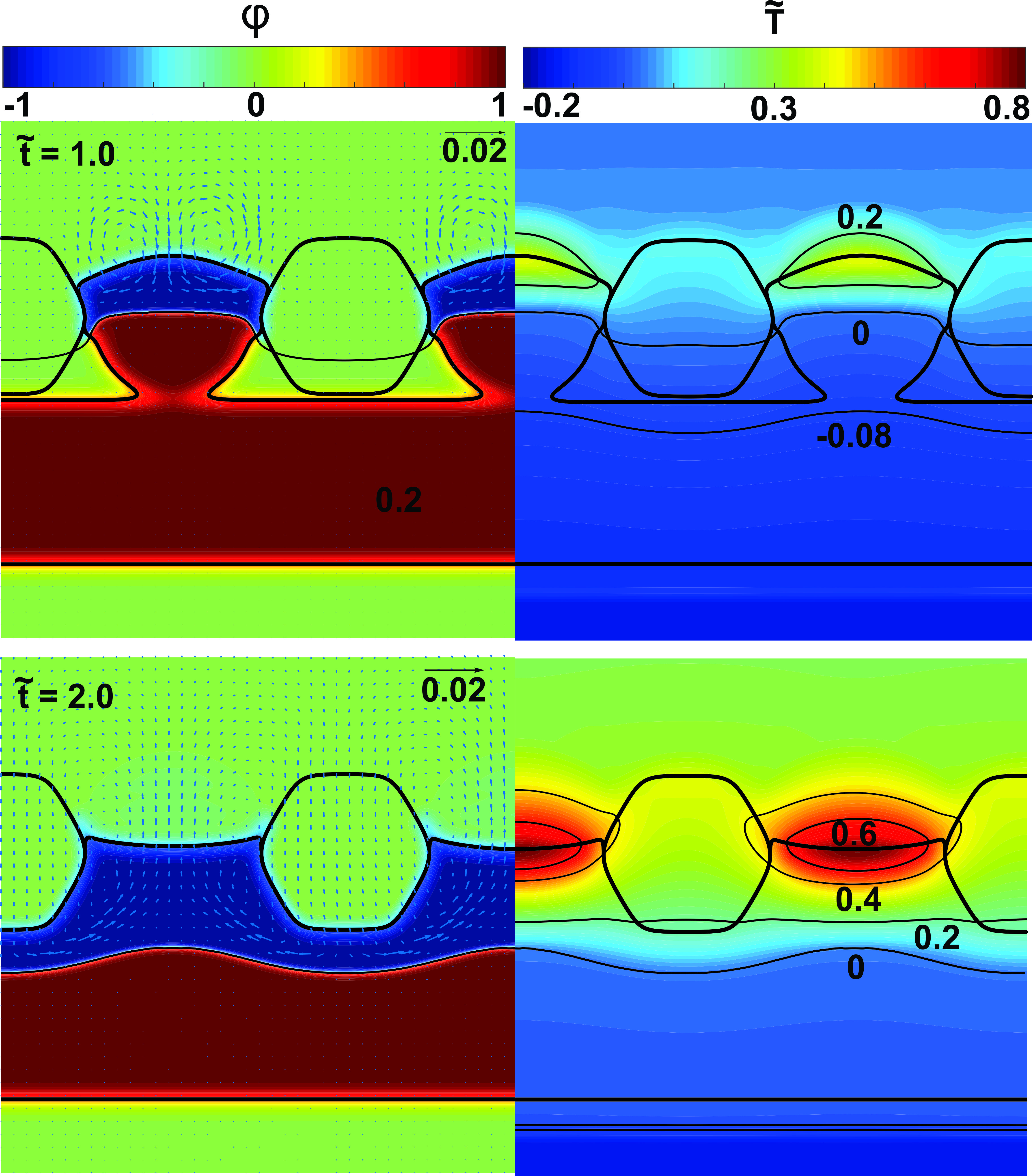}} 
\caption{Velocity field overlapped with phase and temperature fields for the case shown in Fig. 5. The magnitude of velocity vectors has been scaled by characteristic velocity $U=0.73~\rm{m}/\rm{s}$.}
\end{figure}

In Fig. 6, we demonstrate the results by considering the convective effect using the same conditions for the case shown in Fig. 5. The two time instants at $\tilde{t}$ = 1.0 and 2.0 are demonstrated with respect to the same time instants in Fig. 5. A symmetric circulation appears near the top interface of the filler powder due to the tendency of molten filler metal to flatten the free surface on top and wet the diamond grits on the side. Comparing with Fig. 5, the overall interfacial morphology is very similar, indicating that the enhanced wetting due to convective or inertial effect is insignificant under a relatively low laser power. In this case, at around $\mathcal{Q}=2.1\times 10^5~\rm{W}/ \rm{m}$. Away from the coating region of interest, the velocity vanishes in the far field as expected.

\subsection{Brazing process using a scanning laser beam}

\begin{figure*} [h!] \label{f7}
\centerline{\includegraphics[width=\textwidth]{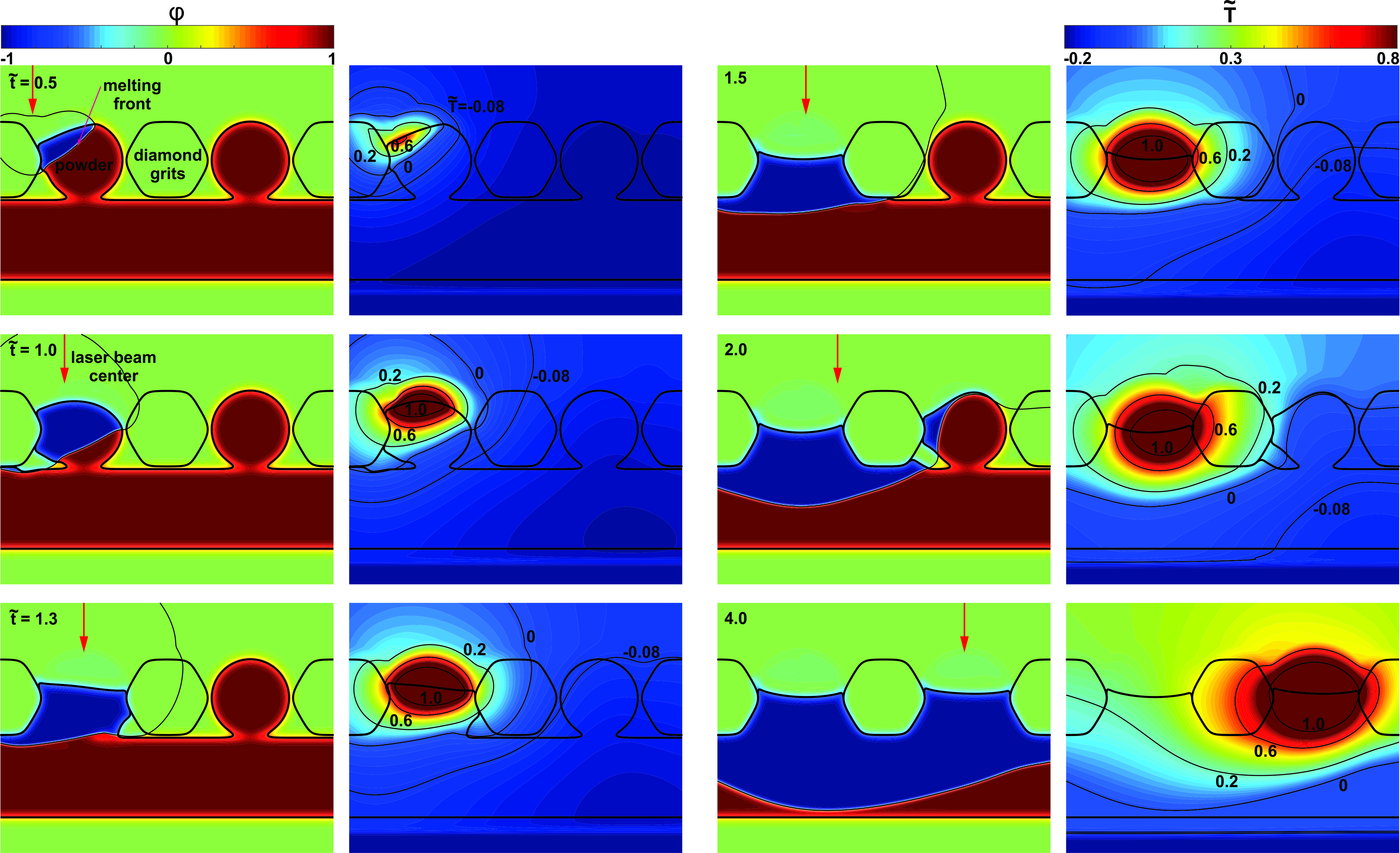}} 
\caption{Transient evolution of phase field $\varphi$ and the scaled temperature field $\tilde{T}$ at six scaled time instants $\tilde{t}$= 0.5, 1.0, 1.3, 1.5, 2.0, and 4.0. The heating dynamics is driven by a scanning laser beam with 2D laser intensity $\mathcal{Q}=2.1\times 10^5~\rm{W}/\rm{m}$, spot size $a=100~\rm{\mu} \rm{m}$, and scanning speed $u=0.1~\textrm{m}/\textrm{s}$. The red arrow indicates the center position of the scanning laser beam.}
\end{figure*}

Figure 7 shows the transient dynamics of the brazing process using an assumed Gaussian laser beam with the same laser power as the case shown in Fig. 5, $\mathcal{Q}=2.1\times 10^5~\textrm{W}/\textrm{m}$, and spot size $a=100~\rm{\mu} \rm{m}$. Although under the same power, the peak value of the heat flux $\pmb H$ from the scanning laser beam is about an order of magnitude higher than the case using a uniform heat flux $\pmb H_u$ . The scanning process starts from the left edge of the computational domain at $x_0=0$ and moves horizontally to the right-hand side with constant speed $u=0.1 ~\textrm{m}/\textrm{s}$. The onset of the melting and wetting appears at the top-left corner around time instant $\tilde{t}=0.5$, where the heating comes from thermal irradiation and heat conduction by direct contact of the filler metal with the diamond grit. The diamond grit has four times higher thermal conductivity and about two times higher thermal diffusivity than the nickel-based filler metal, and thus overall resulting in a smaller temperature gradient (longer thermal diffusive length) in the diamond grits during the heating process. Note that the thermal diffusivities for argon gas, diamond grits, and filler metal are approximately $5.3\times10^{-4}$, $4.5\times10^{-5}$, and $2.1\times10^{-5}~\rm{m}^2/\rm{s}$. Further heating from the scanning beam leads to an evolution of the melting front in the filler powder and the coated filler metal on the substrate. The three-phase contact line moves downward to coat the diamond grit at the left and then moves upward to coat the second diamond grit while filling the gap area ($\tilde{t}=1.0$, 1.3, and 1.5). At time instant $\tilde{t}=1.0$ the molten filler metal shifts to the left and the free surface maintains a circular shape due to strong surface tension effect. The sequential temperature plots clearly show the marching of the melting front at $\tilde{T}$=0 and the accumulation and dissipation of the heat content. With continuous heating, the second filler powder melts and wets more grits ($\tilde{t}=2.0$ and 4.0). The wetting dynamics eventually forms a meniscus between diamond grits.

\begin{figure}[ht]   \label{f8}
\centerline{\includegraphics[width=3.3in]{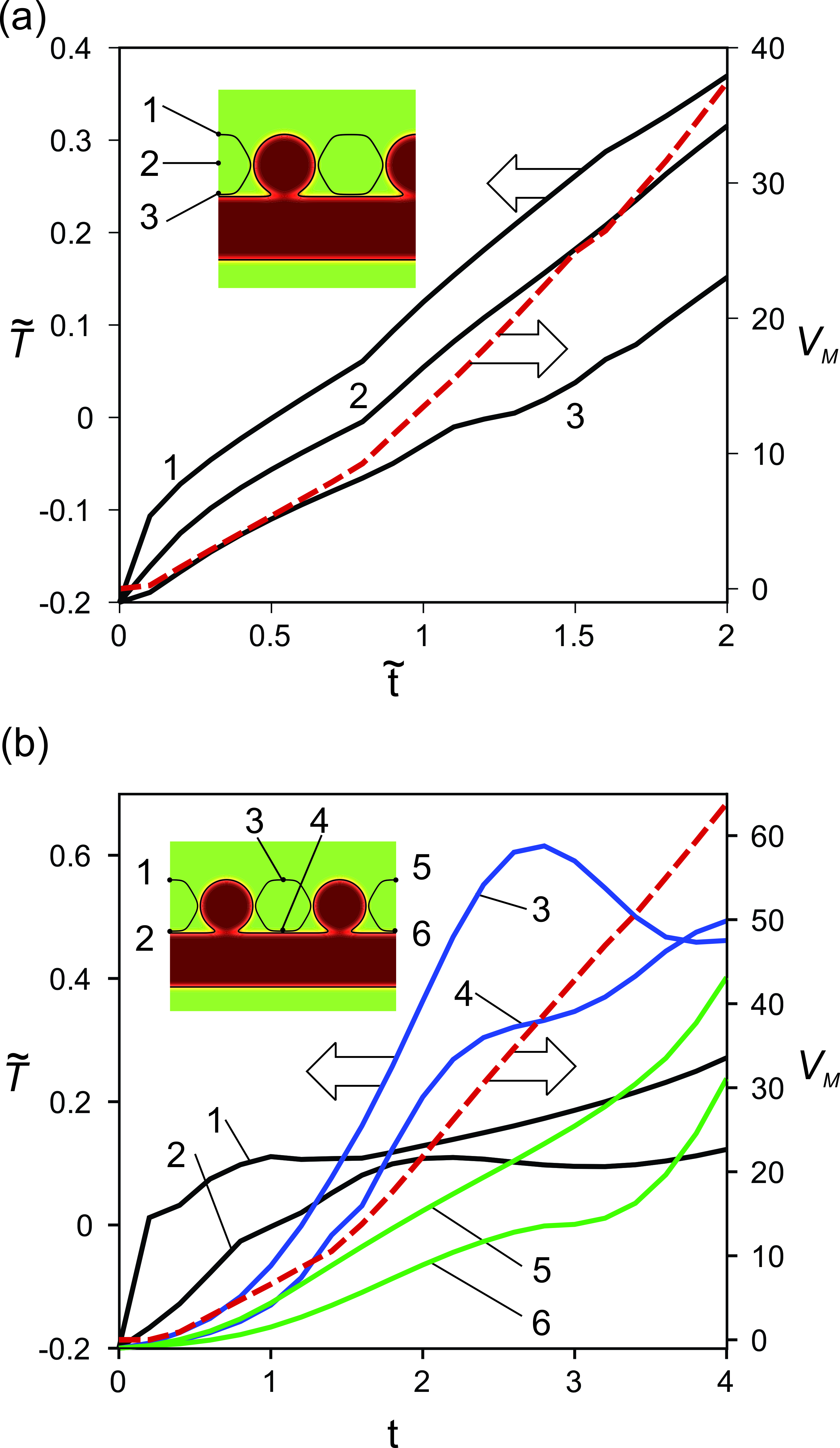}} 
\caption{Temperature history $\tilde{T}(\tilde{t})$ at a few selected points in diamond grits, and the liquid fraction of the overall filler metal $V_\textrm{m}$, either heated by a stationary (a) or scanning (b) laser beam.}
\end{figure}

Figure 8 demonstrates the temperature history of a few points in the diamond grits along with the liquid fraction of the filler metal during the transient process. Shown in Fig. 8a, the process is driven by a stationary laser beam with uniform intensity. The temperature at three selected points within a diamond at the top, middle, and bottom places increases monotonically with first-order type profiles near the initial stage of heating. In the beginning, around $\tilde{t}\simeq0.1$ the top surface reaches the melting temperature $\tilde{T}_\textrm{m}=0$ so that fusion happens and the liquid fraction of the filler metal starts to increase (shown by the red dashed line). At around time $\tilde{t}\simeq0.8$ to 0.9 heating and melting proceed along with capillary wetting which fills the space, resulting in a slightly faster increase of temperature. This is due to the liquid filler coating on diamond grits, which causes higher laser energy absorption ($\alpha_2\simeq0.34$ vs. $\alpha_3\simeq0.2$). In Fig. 8b we demonstrate a few more points to describe the temperature history around the diamond grits as well as the liquid fraction of the filler metal heated by a scanning laser beam. As expected, a strong ramp-up of temperature advances from left to right. Comparing with Fig. 8(a), a much higher temperature gradient and faster temperature rise appear on the surface of the diamond grit. This is due to the focused irradiation near the center point of a Gaussian beam. In the test case, the highest local heat flux is about an order of magnitude higher than the uniform beam. As the scanning proceeds to preheat the second and third diamond grits, at $\tilde{t}\simeq1.0$, the surface temperature of first diamond grit (location 1) decreases due to less thermal irradiation and higher radiation heat loss to the gas environment, shown by temperature increase with a decayed magnitude. Meanwhile, the temperature at the bottom part of the grit continues to raise. The temperature history on location 3 closely correlates with the approaching and departing of the scanning laser beam. The heat conducted through the filler metal around the grit is influenced by the degree of wetting. At $\tilde{t}\simeq2.8$, the highest temperature reaches $\tilde{T}\simeq0.60$, which may cause degradation of the bonding strength due to possible graphitization of the diamond grits. On the other hand, once melting starts from the corner of the powder, the overall liquid fraction of the filler metal increases smoothly and correlates well to the increase of phase transition area. Comparing with the uniform heating case shown in Fig. 8(a), the  demonstrated process using a scanning beam takes longer time ($\tilde{t}=4.0$ vs. $\tilde{t}=2.0$) to complete. However, a larger molten zone is observed in the scanning case ($V_M=68 \%$ vs $37 \%$).

\begin{figure}[ht]  \label{f9}
\centerline{\includegraphics[width=3.3in]{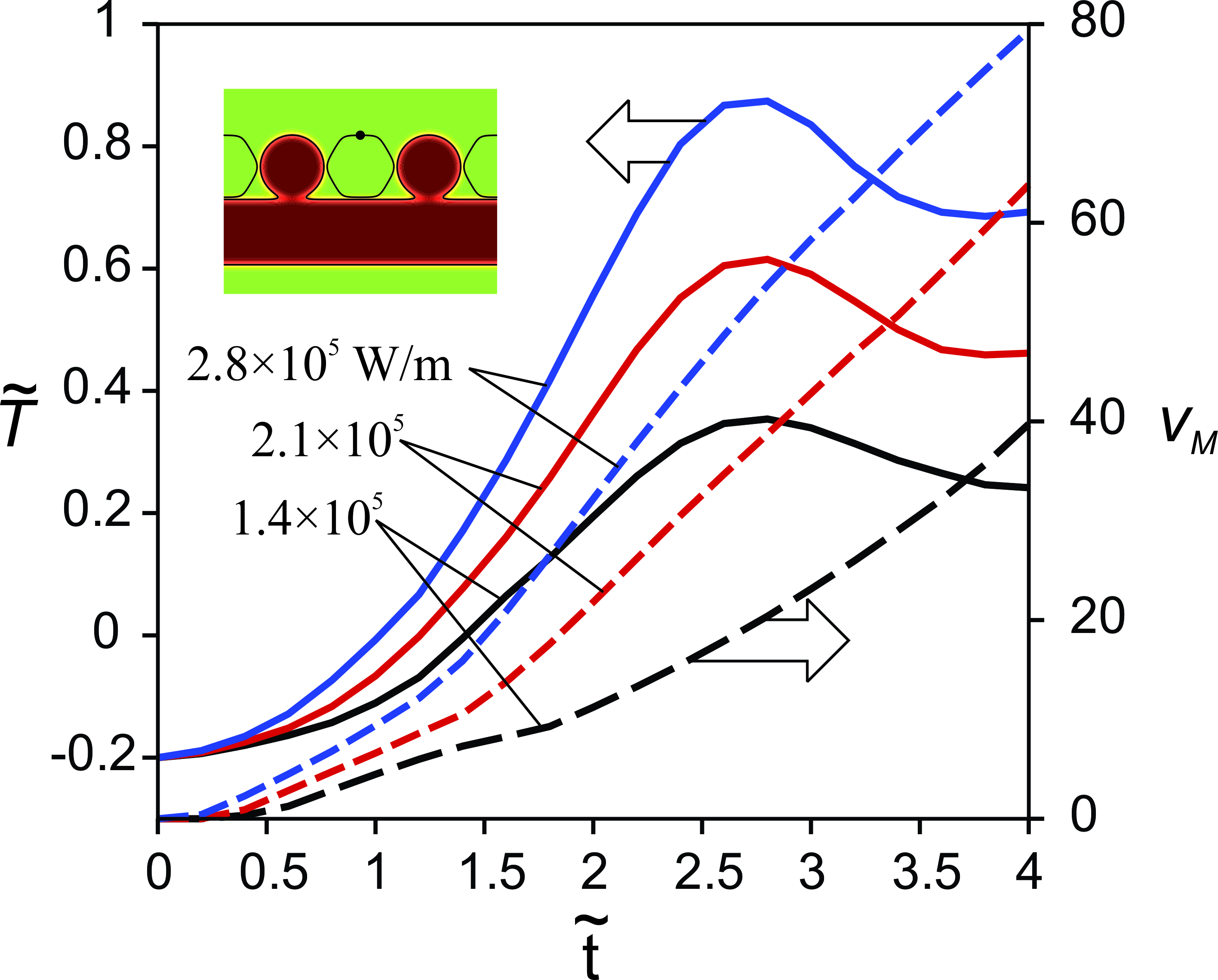}} 
\caption{Temperature history on the top surface of a diamond grit and the corresponding liquid fraction of the filler metal heated by a scanning laser beam with different thermal irradiation power.}
\end{figure}

Figure 9 demonstrates a sensitivity test of brazing dynamics based on different power of thermal irradiation. Applying the same configuration, initial and boundary conditions, and the traveling speed of the laser beam as shown in Fig. 8, with an increasing irradiation intensity $\pmb H$ by adjusting the overall power $Q$. At a higher laser power, the brazing process completes faster but having a risk of overheating the diamond grits even within a short period of time. On the other hand, at lower power, the brazing process may not provide sufficient molten filler metal to bond the diamond grits completely.
 
\newpage
\section{Conclusion}
We present a thermodynamically consistent phase-field model to predict the dynamic process of selective laser brazing of diamond grits. The simplified 2D model features laser heating, melting, and wetting of nickel-based filler metal to diamond grits, as well as filling the voids and providing bonding to the grits. The phase-field framework has successfully incorporated the constraint of three-phase contact line dynamics with a non-isothermal phase transition process as well as the convective effect in the molten liquid metal. Computational results on heating efficiency and transient interfacial evolution under uniform and scanning laser beams are demonstrated. The temperature history, transient liquid fraction, and degree of completeness of the brazing process illustrate the potential of using the theoretical model to predict, design, and optimize the selective laser brazing process at a high level of precision. Future development of the laser brazing model will focus on experimental validation and thermal stress analysis with a different spatial arrangement and diamond protrusion height to facilitate the affinity or bonding strength, and reduce graphitization of diamond grits to enhance cutting tool performance.  
\\\\
\noindent\textbf{Acknowledgments}~~L. Li and T.-H. Fan acknowledge
the financial support of this research from the National Science Foundation (CBET 1930906).

\bibliographystyle{plain}
{}
\end{document}